\documentclass[twocolumn]{aastex631}
\shorttitle{Internal rotation and inclinations}
\shortauthors{Pedersen}
\graphicspath{{./}{figures/}}

\usepackage{amsmath}

\begin{document}

\title{\large Internal rotation and inclinations of slowly pulsating B stars: Evidence of interior angular momentum transport}

\email{mgpedersen@kitp.ucsb.edu}

\author[0000-0002-7950-0061]{May G. Pedersen}
\affiliation{Kavli Institute for Theoretical Physics, Kohn Hall, University of California, 
Santa Barbara, CA 93106, USA}



\begin{abstract}


\noindent One of the largest uncertainties in stellar structure and evolution theory is the transport of angular momentum in the stellar interiors. Asteroseismology offers a powerful tool for measuring the internal rotation frequencies of pulsating stars, but the number of such measurements has remained few for $\gtrsim 3\,{\rm M}_\odot$ main-sequence stars. In this work, we compile a list of 52 slowly pulsating B stars for which the interior rotation has been measured asteroseismically. The measurements of the spin parameters, which describe the relative importance of rotation, for the gravito-inertial mode oscillations show that for 40 of the stars the oscillations fall within the sub-inertial regime. We find that the core rotation frequencies of the stars decrease as a function of age, and show evidence of angular momentum transport occurring on the main-sequence. Finally, we derive the inclination angles of the stars, showing that they are generally consistent with the expectations from surface cancellation effects for the given oscillation modes. 
\end{abstract}



\section{Introduction} \label{sec:intro}


\noindent Asteroseismology, the study and interpretation of stellar oscillations, is not only a powerful tool for studying the internal rotation profiles of stars but also our only means of doing so. The advent of space telescopes providing long-term high-precision photometric light curves has significantly increased the resolution of our measured frequency spectra, allowing us to more easily identify the effects of rotation on the oscillation frequencies. This has opened up a unique avenue to studying angular momentum transport in stars, and has revealed new surprising results, such as the need of much more efficient angular momentum transport to explain the much slower than predicted rotation of the cores of evolved subgiant and red giant stars \citep[e.g.][]{Marques2013,Goupil2013,Cantiello2014,Ouazzani2019}. Current attempts at including additional mechanisms in the theory of angular momentum transport are either still under debate or unable to fully solve this issue \citep{Cantiello2014,Fuller2014,Belkacem2015,Pincon2017,Mathis2018,Fuller2019}.

While the number of stars for which measurements of the internal rotation has increased, the number of intermediate- and high-mass stars on the main-sequence with such measurements has remained small. This is partly caused by the lack of suitable observational data for these stars, but also hindered by the reality that the rotational frequencies of the stars are comparable to the oscillation frequencies, requiring the Coriolis acceleration be taken into account. Recent improvements have been achieved for the $\gamma$~Doradus \citep[$\gamma$~Dor; $1.4\,{\rm M}_\odot \lesssim M \lesssim 2.0\,{\rm M}_\odot$, e.g. ][]{Kaye1999} stars, where the number of stars with measured near-core rotation frequencies has increased from $\approx 50$ stars \citep[e.g.][]{VanReeth2016} to $\geq 600$ \citep{Li2020}. In comparison, the number of main-sequence stars with $\gtrsim 3\,{\rm M}_\odot$ which have such measured near-core rotation frequencies from space photometry were $\leq 10$ \citep[e.g.][]{Aerts2019b}. We refer the reader to \cite{Aerts2019b} and references therein for a detailed review of the current inferences of internal rotation and angular momentum transport from asteroseismology in general, and discuss here the inferred rotation properties of a sample of 52 slowly pulsating B (SPB) stars, for which internal rotation frequencies have been derive asteroseismically. For an overview of asteroseismic inferences obtained for early type stars observed by space telescopes see also \cite{Bowman2020}.

\begin{figure}
\includegraphics[width=\linewidth]{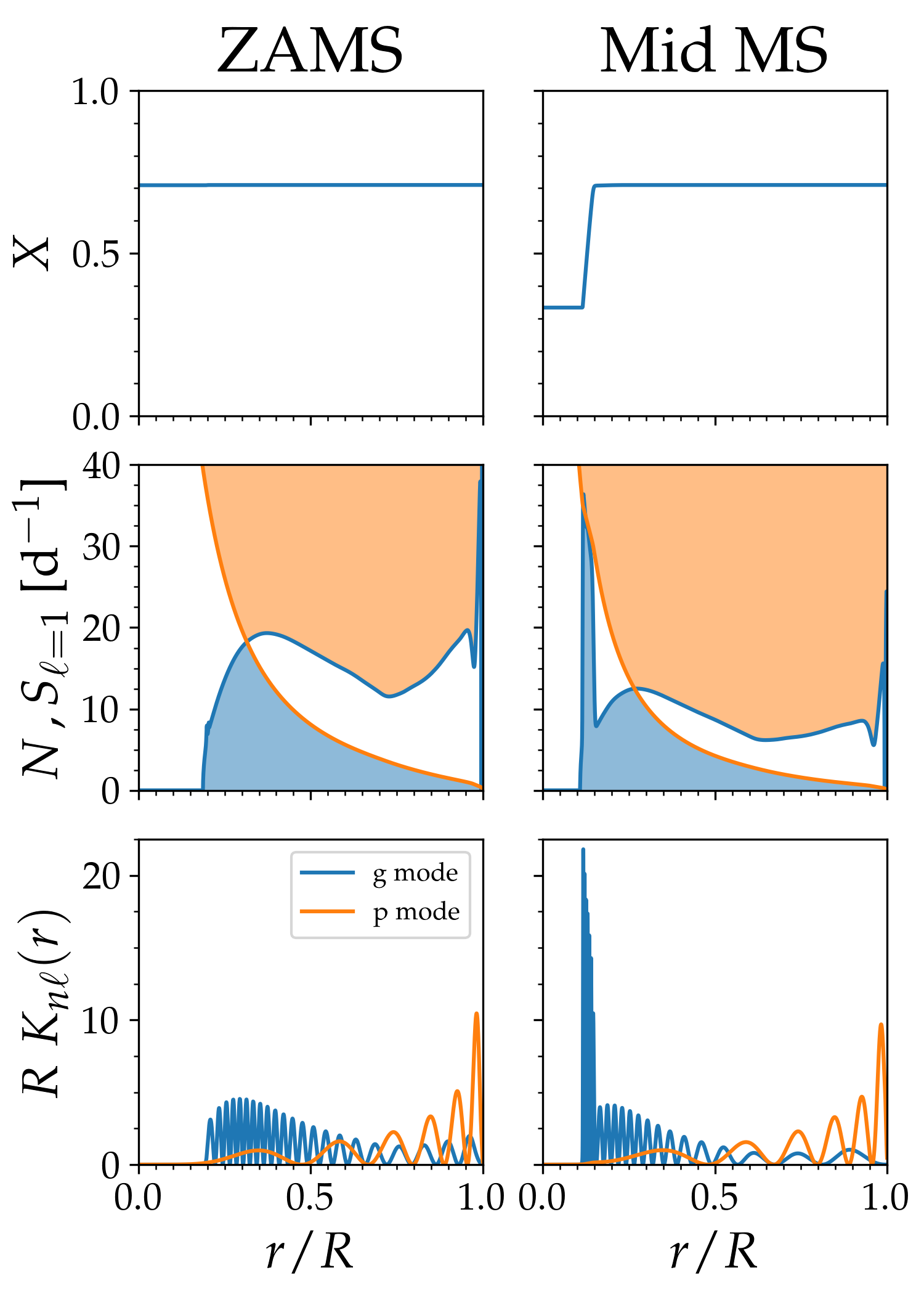}
\caption{Hydrogen mass fraction (X), propagation diagrams, and rotational kernels as a function of fractional radius for a 5\,M$_\odot$ SPB star at two different stages of the main-sequence evolution (left = zero-age main-sequence, right = mid main-sequence). Top: During the main-sequence evolution, the core hydrogen mass fraction has decreased and a chemical gradient been developed which gives rise to the spike in the Brunt-Vaisala frequency shown in blue in the middle right panel. Middle: The blue curve shows the Brunt-Vaisala frequency $N$, while the Lamb frequency for a dipole mode $S_{\ell=1}$ is shown by the orange curve. The blue and orange shaded regions correspond to the propagation cavity of the g- and dipole p-modes, respectively. Bottom: Rotational kernel $K_{n\ell} (r)$ for a dipole prograde g-mode of radial order $n = -22$ (blue), and a $(\ell,m,n)=(1,1,5)$ p-mode (orange) calculated using Eq.~(3.356) in \cite{Aerts2010}. The rotational kernel is made dimensionless by scaling with the stellar radius $R$. The kernels show that the g-modes mainly probe the rotation profile near the convective core, while the p-modes mainly provide information about the rotation near the stellar surface.}\label{fig:gmode_vs_pmode}
\end{figure}

The SPB stars are main-sequence B-type stars ($3\,{\rm M}_\odot \lesssim M \lesssim 10\,{\rm M}_\odot$) which oscillate in non-radial gravity (g) modes, where bouyancy acts as the dominant restoring force \citep[e.g.][]{Aerts2010}. These modes have the highest probing power near the convective cores. Some SPB stars are known to be hybrid pulsators, and also exhibit pressure (p) mode oscillations where pressure acts as the dominant restoring force. Figure~\ref{fig:gmode_vs_pmode} illustrates the difference between the propagation cavities and rotational kernels of p- and g-mode oscillations for a 5\,M$_\odot$ star at two different main-sequence ages. For the remainder of this work we will focus on the g-mode oscillations for the sake of homogeneity of the sample.

The oscillation frequencies are characterized by a set of three quantum numbers: 1) the degree $\ell$ defining the number of surface nodes associated with the oscillation, 2) the azimuthal order $m$ giving the number of the surface nodes crossing the equator, and 3) the radial order $n$ which equals the number of radial nodes of the oscillation. The process of assigning these three numbers to an oscillation is referred to as mode identification. This information is required for asteroseismic modeling, and can be obtained in different ways, some of which also simultaneously provide information about the internal rotation of the stars.

A property of the g-modes is that modes with the same $\ell$ and $m$ values form period spacing patterns, which are the period differences $\Delta P_{\ell, m} = P_{\ell, m, n} - P_{\ell, m, n-1}$ between oscillations consecutive in $n$ plotted as the function of the oscillation period $P_{\ell, m, n}$.
The detection of these patterns is used to identify the modes as required for asteroseismic modeling. For non-rotating, chemically homogeneous stars the period differences are approximately the same across all radial orders and the period spacing pattern is flat. The inclusion of rotation, however, introduces a tilt in the period spacing patterns \citep{Bouabid2013} and the slope of the patterns is used to infer the near-core rotation frequency of the stars \citep{VanReeth2016,Ouazzani2017}. A challenge of using period spacing patterns to simultaneously derive internal rotation frequencies and mode identifications is the high density of g-mode oscillation frequencies in the Fourier spectrum. A high frequency resolution is required in order to identify individual pulsation modes. The high-precision, high-cadence, and long time-base photometric data made available by space telescopes have been crucial for studying period spacing patterns in SPB stars.  A period spacing pattern was detected for the first time for an SPB star \citep[HD~50230][]{Degroote2010} using data from the CoRoT space telescope \citep{Auvergne2009,Baglin2009}.

The rotational frequency of the star can also be measured in two other ways. The first is the case of a slowly rotating star, where the rotational splitting of an oscillation frequency into a $2\ell +1$ multiplet allows one to not only obtain a mode identification of the oscillations, but also measure the rotation frequency based on the width of the splitting. We will return to the subject of rotational frequency splitting in Sect.~\ref{sec:Spin}, and point here the reader to the beautiful example of the lower mass hybrid pulsator KIC~11145123 \citep[][see e.g. their figure 3]{Kurtz2014}. As in the case of the period spacing pattern, a high-frequency resolution is required in order to detect rotationally split oscillation modes. Finally, in the absence of period spacing patterns and rotational split oscillation modes, as is generally the case for SPB stars with ground-based data only, one may attempt to derive the internal rotation frequency when carrying out mode-identification from multi-color photometry. Assuming uniform rotation,
\citet{DaszynskaDaszkiewicz2007,DaszynskaDaszkiewicz2008,DaszynskaDaszkiewicz2015} demonstrated that including the effect of rotation in the calculation of the amplitude ratios from different photometric passbands allows one to simultaneously derive $\ell, m$, and the rotational velocity of the star.

In this work we compile a list of 52 SPB stars for which their interior rotation has been measured with asteroseismology using data from space telescopes or multi-color ground-based observing campaigns. An overview of the sample is provided in Sect.~\ref{sec:Sample}. We start by calculating the rotational spin parameters in Sect.~\ref{sec:Spin}, and investigate the age dependence of the near core rotation in Sect.~\ref{Sec:Rot_vs_Age}. 
We look for evidence of angular momentum transport in Sect.~\ref{Sec:AM}. The assumption of rigid rotation is then used to infer the inclination angles and its relation to the mode identification of the oscillation frequencies in Sect.~\ref{Sec:Inclination}. Finally, our conclusions are presented in Sect.~\ref{sec:Conclusions}.

\section{Overview of the SPB star sample}\label{sec:Sample}

\begin{figure}
\includegraphics[width=\linewidth]{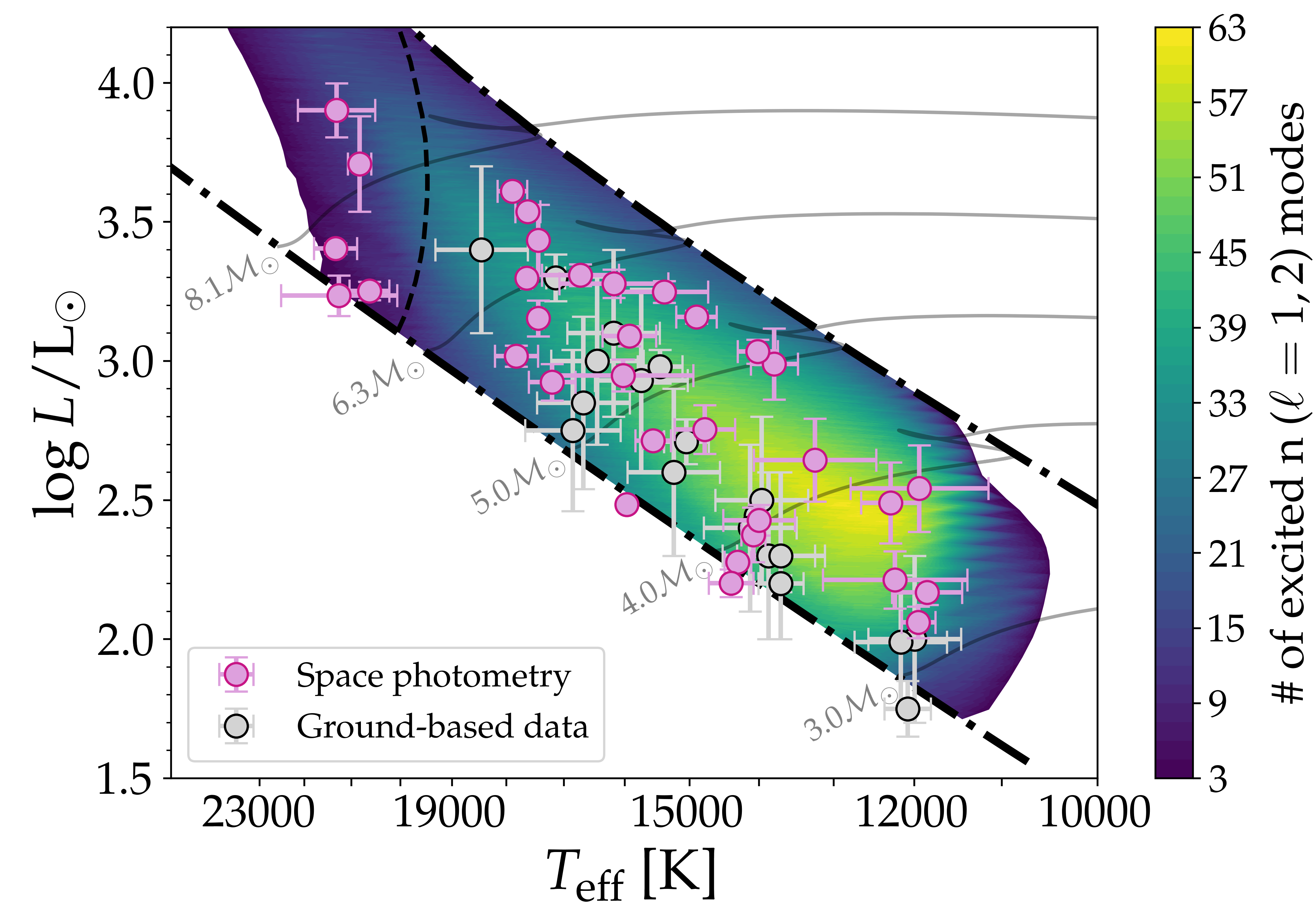}
\caption{SPB instability strip and the locations of the sample of SPB stars studied in this work for which derived luminosities are available. The shaded region indicates the number of  radial orders of degree $\ell = 1,2$ that are expected to be excited in a given region of the HR diagram \citep{Moravveji2016}. Example evolutionary tracks for five initial stellar masses are shown in grey. The dot-dashed lines indicate the boundaries of the main-sequence, and the dashed black line shows the beginning of the $\beta$ Cephei instability strip. The colors of the data points indicate if the rotational frequency was derived based on space photometry (pink) or ground-based data (grey).}\label{fig:HRD}
\end{figure}

The list of 52 SPB stars considered in this work was constructed by carrying out a literature search to identify known SPB stars, which have 1) an interior rotation estimate derived using their g-mode oscillations, and 2) an unambiguous mode identification for at least one of their g-mode oscillation frequencies. The sample of stars is shown in the HR diagram in Fig.~\ref{fig:HRD}. We group the stars in two separate samples, one for which the rotation frequency measurements and mode identification was done based on space photometry (pink) and the other where only ground-based data was available (grey). We discuss each of the two samples separately below.
 
For all of the stars in the sample, we likewise collect three different types of age indicators, which are used in Sect.~~\ref{Sec:Rot_vs_Age} to investigate the main-sequence evolution of the internal rotation. For the first age indicator we use the ratio $X_{\rm c}/X_{\rm ini}$ of the current core hydrogen mass fraction $X_{\rm c}$ to the initial hydrogen mass fraction $X_{\rm ini}$. These ratios are generally available in the cases where an asteroseismic modeling has been carried out for the star. Secondly, we use the ratio $t/t_{\rm MS}$ between the current age of the star $t$ and the main-sequence lifetime $t_{\rm MS}$. These literature values were estimated by comparing the position of the stars in the HR diagram to stellar evolution tracks, and are generally available for the SPB stars with ground-based data. Finally, the surface gravity $\log g$ is used as an age indicator. For six of the stars in the sample, this was the only available age indicator. A comparison between $X_{\rm c}/X_{\rm ini}$, $t/t_{\rm MS}$, and $\log g$ is given in Fig.~\ref{fig:age_ind} in Appendix~\ref{App:Age_indicators}. For stars where either $X_{\rm c}/X_{\rm ini}$ or $t/t_{\rm MS}$ values are available, the same figure is used to make a conversion between the two, see Appendix~\ref{App:Age_indicators} for details. For stars where stellar parameters and rotation rates have been derived but no errors on these estimates were provided, we adopt the median fractional errors\footnote{As an example, the error on the rotational frequency $\sigma_{f_{\rm rot}}$ for an SPB star where $\sigma_{f_{\rm rot}}$ has not been estimated is taken to be $\sigma_{f_{\rm rot}}^\star = f_{\rm rot}^\star \times (\text{median}\,\sigma_{f_{\rm rot}}/f_{\rm rot}\text{ of the sample})$.} of either the sample with space photometry or ground-based data. An overview of the sample is given in Table~\ref{Tab:SPB_sample}.

As discussed in Appendix~\ref{App:Age_indicators}, both $X_{\rm c}/X_{\rm ini}$ and $t/t_{\rm MS}$ are better age indicators than $\log g$ as these ratios are less affected by the mass and by the internal mixing history of the stars. For the same reason, we recommend that future asteroseismic modeling efforts include both ratios as output of the modeling rather than just $X_{\rm c}$ and $t$ values.

\subsection{Space photometry}\label{sec:SPB_space}

33 of the 52 SPB stars were observed by space telescopes and 31 of these have undergone asteroseismic modeling. The largest sample was taken from \cite{Pedersen2021}, who recently carried out an asteroseismic modelling of the dipole period spacing patterns in 26 SPB stars observed by the \emph{Kepler} space telescope \citep{Koch2010,Borucki2010}. For 11 of these stars, patterns had previously been identified \citep{Papics2014,Papics2015,Papics2017,Szewczuk2018,Zhang2018} but only three of them had been asteroseismically modelled \citep{Moravveji2015,Moravveji2016a,Szewczuk2018,Michielsen2021,Bowman2021}.  \cite{Pedersen2021} discovered period spacing patterns in 15 new SPB stars and performed asteroseismic modelling of the entire sample of 26 SPB stars. In this work, we adopt the updated parameter estimates from \cite{Pedersen2022}, which take into account that different shapes of the internal mixing profiles do equally well at explaining the observed period spacing patterns for some of the stars\footnote{The parameter values from Table~4 and 5 in \cite{Pedersen2022} were used in this paper. For the five stars for which a single shape of the internal mixing profile was found to explain the observed period spacing patterns the best, the values from Tables~2 and 3 in \cite{Pedersen2022} were used instead.}.

In addition to these 26 stars, two more SPB stars observed by \emph{Kepler} (KIC~8264293 and KIC~8324483) have period spacing patterns which have been modeled asteroseismically. The period spacing pattern of KIC~8264293 was identified by \cite{Szewczuk2021}, and the subsequent asteroseismic modeling carried out by \cite{Szewczuk2022} revealed that the star is close to the zero-age main-sequence. \cite{Zhang2018} identified a period spacing pattern in KIC~8324483. The initial modeling of the pattern \citep{Deng2018} resulted in a mode identification of $(\ell,m) = (1,0)$ and was later improved by \cite{Wu2020}, revealing that the star is $\approx 33\%$ through its main-sequence evolution.

Two of the stars in the sample with space photometry (HD~50230 and HD~43317) were observed by the CoRoT space telescope. \cite{Degroote2010} identified a period spacing pattern in HD~50230 consisting of dipole zonal modes $(\ell,m) = (1,0)$. The star was later identified to be a spectroscopic binary, exhibiting both g-mode pulsations and pressure (p) modes \citep{Degroote2012}, the latter of which showing a frequency splitting $\Delta f_{\rm p} = 0.044\pm 0.007$\,d$^{-1}$ likely caused by rotation. The authenticity of the pattern has later been questioned by \cite{Szewczuk2015b}. Using the observed frequency splitting of the p-modes, \cite{Wu2019} derived the rotational frequency of the star to be $f_{\rm rot} = 0.007\pm 0.001$\,d$^{-1}$ and carried out an asteroseismic modeling of the star based on the g-mode period spacing pattern. Fitting the slope of the period spacing pattern following the method by \cite{VanReeth2016} results in a rotation frequency of $0.0955^{+0.0955}_{-0.0637}$\,d$^{-1}$ assuming $(\ell,m) = (1,0)$. 

HD~43317 is a magnetic \citep{Briquet2013,Buysschaert2017} B-type star, and is one out of two stars in the sample with space photometry without an identified period spacing pattern. The star was initially thought to be a hybrid pulsator, with p- and g-mode oscillations detected in the CoRoT data of HD~43317 by \cite{Papics2012}, who identified one of the g-mode oscillations to be a prograde quadropole $(\ell, m) = (2, 2)$ mode based on spectroscopic mode identification. \cite{Buysschaert2018} reanalyzed the CoRoT data of the star and carried out an asteroseismic modelling, in which they identified ten additional $(\ell,m) = (1,-1)$ g-modes and five $(\ell,m) = (2,-1)$ g-modes and found the star to be an SPB star rather than a hybrid pulsator\footnote{The previously identified p-modes turned out to be rotationally shifted g-modes.}. As the only exception to the rest of the 51 SPB stars, the rotation frequency of HD~43317 was derived from a magnetometric analysis of the star rather than from asteroseismology \citep{Buysschaert2017}, and then adopted for the asteroseismic modeling of the star assuming rigid rotation.

One star (HD~201433) in the sample has been asteroseismically modeled and had its internal rotation profile derived based on data from the BRITE satellite constellation \citep{Weiss2014}. HD~201433 is a triple system with one short period and one long period binary component, and has been studied in detail by \cite{Kallinger2017} who also carried out an asteroseismic modeling of the star. Aside from KIC~10526294, it is the only star in the sample rotating slowly enough to exhibit clear rotationally split dipole g-modes, which were used to infer the internal rotation profile of the star, indicating that its radiative envelope is rigidly rotating up to $\sim 90\%$ of the stellar radius.

The last two stars (HD~261810 and HD~48012) in the sample of SPB stars with space photometry were observed by the MOST space telescope \citep{Walker2003} and lie in the field of the open cluster NGC~2264. \cite{Zwintz2017} identified a period spacing pattern in both stars based on the data from MOST, which were asteroseismically modelled to show that both stars are on or just approaching the zero-age main-sequence (ZAMS). For HD~261810, \cite{Zwintz2017} found the period spacing pattern to consists of either $(\ell,m) = (1,1)$ or $(2,0)$ modes, meaning that the mode identification was not determined unambiguously. We nevertheless choose to include the star in our sample, as it is one on only few young SPB stars with a rotational frequency estimate available. For remainder of this work, we will assume that the pattern consists of $(\ell,m) = (1,1)$ modes.

\subsection{Ground-based data}

For the remaining 19 out of our sample of 52 SPB stars, the internal rotation frequency was derived based on data from ground-based observing campaigns. \cite{Szewczuk2015a} performed mode identification and estimated rotational velocities assuming uniform rotation of 31~SPB stars for which multi-colour UBV Geneva photometry was available from ground-based observing campaigns \citep[using the oscillation frequencies and amplitudes provided by][]{North1994,DeCat2002,DeCat2002b,DeCat2007,Szewczuk2015a}. For 16 of the 31 SPB stars, \cite{Szewczuk2015a} obtained an unambiguously mode identification. These 16 stars have therefore been included in our sample of SPB stars. Asteroseismic modeling was previously  achieved for one of the stars, HD~21071, but unfortunately the main-sequence age was not provided \citep{Szewczuk2015b}.

For the remaining three stars in the sample (HD~36999, HD~66181, and HD~97895), the internal rotation was derived by combining photometric data from five different ground based surveys \citep{Fedurco2020}. The mode identification of their oscillation frequencies was obtained by comparing the observed oscillation frequencies to a grid of theoretical oscillation frequencies computed under the assumption that the stars are slowly rotating. While this modeling procedure also provided an absolute age estimate for each of the stars, a fractional main-sequence age is not available. 

We note that in this sample of 19 SPB stars there is a lack of fast-rotators. All but two stars have projected equatorial velocities below 60\,km\,s$^{-s}$, while the fastest rotating star in the sample rotates more than twice as slow than the fastest rotating star in the sample of SPB stars with space photometry, see Table~\ref{Tab:rot_parameters}. This lack of fast rotators was also previously noted by \cite{DeCat2002} and \cite{Szewczuk2015a}.

{\movetabledown=2.05in\tabcolsep=5pt
\startlongtable
\begin{deluxetable*}{lcccccccc}
\tablecaption{Overview of the sample of 52 SPB stars and their three different age indicators.\label{Tab:SPB_sample}}
\tablewidth{700pt}
\tabletypesize{\small}
\tablehead{
\colhead{Star ID} & 
\colhead{Data source} & 
\colhead{Known binary} & 
\colhead{$M$} & 
 \colhead{$X_{\rm c}/ X_{\rm ini}$} &
 \colhead{$t/ t_{\rm MS}$} &
  \colhead{$\log g$} &
\colhead{Reference}\\
& & &  \colhead{(M$_\odot$)} &  \colhead{(\%)} & \colhead{(\%)}
} 
\startdata
		 KIC 3240411 	&	 Kepler 	&	 N 	&	5.34$^{+0.04}_{-0.04}$	&	 68.33$^{+0.23}_{-0.23}$ 	&	 51.56$\pm 0.28$ 	&	 4.09$^{+0.05}_{-0.02}$ 	&	 1\\[0.5ex]
		 KIC 3459297 	&	 Kepler 	&	 N	&	 3.63$^{+0.66}_{-0.66}$ 	&	 17.98$^{+13.94}_{-13.94}$ 	&	 92.66$\pm 7.16$ 	&	 3.79$^{+0.12}_{-0.12}$ 	&	 1\\[0.5ex]
		 KIC 3865742 	&	 Kepler 	&	 N 	&	 5.81$^{+0.15}_{-0.15}$ 	&	 28.13$^{+6.19}_{-6.19}$ 	&	 86.98$\pm 3.80$ 	&	 3.69$^{+0.13}_{-0.13}$ 	&	 1\\[0.5ex]
		 KIC 4930889 	&	 Kepler 	&	 Y 	&	 4.06$^{+0.31}_{-0.31}$	&	 36.21$^{+0.73}_{-0.73}$ 	&	 81.67$\pm 0.52$ 	&	 3.91$^{+0.00}_{-0.00}$ 	&	 1\\[0.5ex]
		 KIC 4936089 	&	 Kepler 	&	 N 	&	 3.76$^{+0.24}_{-0.24}$ 	&	 36.55$^{+13.41}_{-13.41}$ 	&	 81.42$\pm 9.58$ 	&	 3.83$^{+0.09}_{-0.09}$ 	&	 1\\[0.5ex]
		 KIC 4939281 	&	 Kepler 	&	 N 	&	 5.38$^{+0.01}_{-0.01}$ 	&	 48.01$^{+0.09}_{-0.09}$ 	&	 72.45$\pm 0.08$ 	&	 3.70$^{+0.05}_{-0.03}$ 	&	 1\\[0.5ex]
		 KIC 5309849 	&	 Kepler 	&	 N 	&	 5.67$^{+0.07}_{-0.07}$ 	&	 15.42$^{+0.04}_{-0.04}$ 	&	 93.93$\pm 0.02$ 	&	 3.70$^{+0.05}_{-0.05}$ 	&	 1\\[0.5ex]
		 KIC 5941844 	&	 Kepler 	&	 N 	&	 3.66$^{+0.09}_{-0.09}$ 	&	 93.99$^{+0.44}_{-0.44}$ 	&	 12.62$\pm 0.82$ 	&	 4.30$^{+0.02}_{-0.02}$ 	&	 1\\[0.5ex]
		 KIC 6352430 	&	 Kepler 	&	 Y 	&	 3.18$^{+0.23}_{-0.23}$ 	&	 53.23$^{+24.90}_{-24.90}$ 	&	 67.75$\pm 24.13$ 	&	 4.04$^{+0.15}_{-0.15}$ 	&	 1\\[0.5ex]
		 KIC 6462033 	&	 Kepler 	&	 N 	&	 7.27$^{+0.17}_{-0.17}$ 	&	 14.92$^{+1.86}_{-1.86}$ 	&	 94.17$\pm 0.90$ 	&	 3.67$^{+0.03}_{-0.03}$ 	&	 1\\[0.5ex]
		 KIC 6780397 	&	 Kepler 	&	 N 	&	 4.87$^{+1.08}_{-1.08}$ 	&	 28.44$^{+23.80}_{-23.80}$ 	&	 86.78$\pm 14.99$ 	&	 3.64$^{+0.08}_{-0.08}$ 	&	 1\\[0.5ex]
		 KIC 7630417 	&	 Kepler 	&	 N 	&	 7.01$^{+0.09}_{-0.09}$ 	&	 15.33$^{+0.25}_{-0.25}$ 	&	 93.97$\pm 0.12$ 	&	 3.68$^{+0.01}_{-0.01}$ 	&	 1\\[0.5ex]
		 KIC 7760680 	&	 Kepler 	&	 N 	&	 3.43$^{+0.06}_{-0.06}$ 	&	 60.63$^{+7.94}_{-7.94}$ 	&	 60.34$\pm 8.50$ 	&	 4.01$^{+0.02}_{-0.02}$ 	&	 1\\[0.5ex]
		 KIC 8057661 	&	 Kepler 	&	 N 	&	 9.14$^{+0.59}_{-0.59}$ 	&	 17.11$^{+1.69}_{-1.69}$ 	&	 93.10$\pm 0.85$ 	&	 3.77$^{+0.00}_{-0.00}$ 	&	 1\\[0.5ex]
		 KIC 8255796 	&	 Kepler 	&	 N 	&	 5.73$^{+0.01}_{-0.01}$ 	&	 2.01$^{+0.04}_{-0.04}$ 	&	 99.64$\pm 0.02$ 	&	 3.64$^{+0.05}_{-0.03}$ 	&	 1\\[0.5ex]
		 KIC 8264293 	&	 Kepler 	&	 N 	&	 3.54$^{+0.06}_{-0.12}$ 	&	 98.59$^{+1.41}_{-3.11}$ 	&	 3.73$\pm 6.97$ 	&	 4.37$^{+0.05}_{-0.06}$ 	&	 2\\[0.5ex]
		 KIC 8324482 	&	 Kepler 	&	 N 	&	 6.08$^{+0.14}_{-0.25}$ 	&	 71.07$^{+0.25}_{-1.04}$ 	&	 48.14$\pm 0.82$ 	&	 4.25$^{+0.03}_{-0.04}$ 	&	 3\\[0.5ex]
		 KIC 8381949 	&	 Kepler 	&	 N 	&	 7.69$^{+1.06}_{-1.06}$ 	&	 43.35$^{+9.34}_{-9.34}$ 	&	 76.32$\pm 7.46$ 	&	 3.84$^{+0.09}_{-0.09}$ 	&	 1\\[0.5ex]
		 KIC 8459899 	&	 Kepler 	&	 Y 	&	 3.94$^{+0.61}_{-0.61}$ 	&	 26.38$^{+7.89}_{-7.89}$ 	&	 88.03$\pm 4.69$ 	&	 3.86$^{+0.10}_{-0.10}$ 	&	 1\\[0.5ex]
		 KIC 8714886 	&	 Kepler 	&	 N 	&	 6.09$^{+0.24}_{-0.24}$ 	&	 29.56$^{+0.37}_{-0.37}$ 	&	 86.09$\pm 0.23$ 	&	 3.86$^{+0.01}_{-0.01}$ 	&	 1\\[0.5ex]
		 KIC 8766405 	&	 Kepler 	&	 N 	&	 4.40$^{+0.48}_{-0.48}$ 	&	 12.58$^{+6.52}_{-6.52}$ 	&	 95.27$\pm 3.01$ 	&	 3.59$^{+0.07}_{-0.07}$ 	&	 1\\[0.5ex]
		 KIC 9020774 	&	 Kepler 	&	 N 	&	 3.47$^{+0.04}_{-0.04}$ 	&	 60.20$^{+15.63}_{-15.63}$ 	&	 60.79$\pm 16.75$ 	&	 4.15$^{+0.08}_{-0.08}$ 	&	 1\\[0.5ex]
		 KIC 9715425 	&	 Kepler 	&	 N 	&	 4.66$^{+0.38}_{-0.38}$ 	&	 32.11$^{+13.61}_{-13.61}$ 	&	 84.45$\pm 9.00$ 	&	 3.77$^{+0.06}_{-0.06}$ 	&	 1\\[0.5ex]
		 KIC 10526294 	&	 Kepler 	&	 N 	&	 3.64$^{+0.05}_{-0.15}$ 	&	 23.29$^{+3.95}_{-3.95}$ 	&	 89.81$\pm 2.22$ 	&	 3.74$^{+0.13}_{-0.07}$ 	&	 1\\[0.5ex]
		 KIC 10536147 	&	 Kepler 	&	 N 	&	 6.51$^{+0.51}_{-0.51}$ 	&	 74.28$^{+11.73}_{-11.73}$ 	&	 43.93$\pm 15.86$ 	&	 4.11$^{+0.07}_{-0.07}$ 	&	 1\\[0.5ex]
		 KIC 11360704 	&	 Kepler 	&	 N 	&	 5.13$^{+0.75}_{-0.75}$ 	&	 45.13$^{+10.78}_{-10.78}$ 	&	 74.88$\pm 8.89$ 	&	 3.90$^{+0.09}_{-0.09}$ 	&	 1\\[0.5ex]
		 KIC 11971405 	&	 Kepler 	&	 N 	&	 3.59$^{+0.08}_{-0.08}$ 	&	 44.13$^{+2.31}_{-2.31}$ 	&	 75.70$\pm 1.87$ 	&	 3.99$^{+0.00}_{-0.00}$ 	&	 1\\[0.5ex]
		 KIC 12258330 	&	 Kepler 	&	 N 	&	 3.62$^{+0.09}_{-0.09}$ 	&	 74.28$^{+6.62}_{-6.62}$ 	&	 43.93$\pm 8.91$ 	&	 4.28$^{+0.02}_{-0.02}$ 	&	 1\\[0.5ex]
		 HD 43317 	&	 CoRoT 	&	 N 	&	 5.80$^{+0.10}_{-0.20}$ 	&	 77.14$^{+1.43}_{-2.86}$ 	&	 40.00$\pm 2.98$ 	&	 4.14$^{+0.01}_{-0.02}$ 	&	 4\\[0.5ex]
		 HD 50230 	&	 CoRoT 	&	 Y 	&	 6.21$^{+0.06}_{-0.06}$ 	&	 43.10$^{+1.41}_{-1.13}$ 	&	 76.52$\pm 1.01$ 	&	 3.72$^{+0.00}_{-0.01}$ 	&	 5\\[0.5ex]
		 HD 48012 	&	 MOST 	&	 N 	&	 4.00$^{+0.07}_{-0.14}$ 	&	 100.00$^{+0.00}_{-3.16}$ 	&	 0.88$\pm 5.75$ 	&	 4.30$^{+0.15}_{-0.15}$ 	&	6\\[0.5ex]
		 HD 261810 	&	 MOST 	&	 N 	&	 5.50$^{+0.09}_{-0.19}$ 	&	 100.00$^{+0.00}_{-3.16}$ 	&	 0.88$\pm 5.75$ 	&	 4.25$^{+0.10}_{-0.10}$ 	&	 6\\[0.5ex]
		 HD 201433 	&	 BRITE 	&	 Y 	&	 3.05$^{+0.05}_{-0.11}$ 	&	 57.29$^{+9.00}_{-9.00}$ 	&	 63.80$\pm 9.11$ 	&	 4.15$^{+0.07}_{-0.07}$ 	&	 7\\[0.5ex]
		 HD 1976 	&	 ground based 	&	 Y 	&	 5.10$^{+1.00}_{-1.00}$ 	&	 43.75$^{+28.37}_{-28.37}$ 	&	 76.00$\pm 21.00$ 	&	 3.97$^{+0.20}_{-0.20}$ 	&	 8\\[0.5ex]
		 HD 21071 	&	 ground based 	&	 N 	&	 4.00$^{+0.60}_{-0.60}$ 	&	 77.85$^{+22.01}_{-22.01}$ 	&	 39.00$\pm 30.00$ 	&	 4.09$^{+0.05}_{-0.05}$ 	&	 8, 9\\[0.5ex]
		 HD 24587 	&	 ground based 	&	 Y 	&	 3.70$^{+0.50}_{-0.50}$ 	&	 79.92$^{+20.43}_{-20.43}$ 	&	 36.00$\pm 29.00$ 	&	 4.26$^{+0.20}_{-0.20}$ 	&	 10\\[0.5ex]
		 HD 25558 	&	 ground based 	&	 Y 	&	 \dots 	&	 \dots 	&	 \dots 	&	 4.20$^{+0.20}_{-0.20}$ 	&	 11\\[0.5ex]
		 HD 26326 	&	 ground based 	&	 N 	&	 4.50$^{+0.70}_{-0.70}$ 	&	 70.39$^{+27.23}_{-27.23}$ 	&	 49.00$\pm 32.00$ 	&	 4.14$^{+0.20}_{-0.20}$ 	&	 11\\[0.5ex]
		 HD 36999 	&	 ground based 	&	 N 	&	 3.60$^{+0.06}_{-0.12}$&	 \dots 	&	 \dots	&	 4.31$^{+0.05}_{-0.06}$ 	&	 12\\[0.5ex]
		 HD 53921 	&	 ground based 	&	 Y 	&	 3.70$^{+0.50}_{-0.50}$ 	&	 77.85$^{+22.01}_{-22.01}$ 	&	 39.00$\pm 30.00$ 	&	 4.23$^{+0.20}_{-0.20}$ 	&	 10\\[0.5ex]
		 HD 66181 	&	 ground based 	&	 N 	&	 \dots 	&	 \dots	&	 \dots 	&	 4.35$^{+0.05}_{-0.06}$ 	&	 12\\[0.5ex]
		 HD 74195 	&	 ground based 	&	 N 	&	 5.50$^{+1.00}_{-1.00}$ 	&	 35.73$^{+28.04}_{-28.04}$ 	&	 82.00$\pm 18.00$ 	&	 3.91$^{+0.20}_{-0.20}$ 	&	 10\\[0.5ex]
		 HD 85953 	&	 ground based 	&	 N 	&	 6.80$^{+1.40}_{-1.40}$ 	&	 35.73$^{+28.04}_{-28.04}$ 	&	 82.00$\pm 18.00$ 	&	 3.91$^{+0.20}_{-0.20}$ 	&	 13\\[0.5ex]
		 HD 92287 	&	 ground based 	&	 Y 	&	 5.40$^{+1.00}_{-1.00}$ 	&	 50.80$^{+27.33}_{-27.33}$ 	&	 70.00$\pm 23.00$ 	&	 4.00$^{+0.20}_{-0.20}$ 	&	 10\\[0.5ex]
		 HD 97895 	&	 ground based 	&	 N 	&	 4.80$^{+0.08}_{-0.17}$ 	&	 \dots 	&	 \dots	&	 4.39$^{+0.05}_{-0.06}$ 	&	 12\\[0.5ex]
		 HD 121190 	&	 ground based 	&	 N 	&	 2.80$^{+0.30}_{-0.30}$ 	&	 \dots	&	 \dots 	&	 4.43$^{+0.09}_{-0.09}$ 	&	 13\\[0.5ex]
		 HD 123515 	&	 ground based 	&	 Y 	&	 3.00$^{+0.40}_{-0.40}$ 	&	 79.92$^{+19.67}_{-19.67}$ 	&	 36.00$\pm 28.00$ 	&	 4.27$^{+0.20}_{-0.20}$ 	&	 10\\[0.5ex]
		 HD 160762 	&	 ground based 	&	 Y 	&	 6.14$^{+0.22}_{-0.22}$ 	&	 \dots 	&	 \dots 	&	 3.82$^{+0.07}_{-0.07}$ 	&	 14\\[0.5ex]
		 HD 179588 	&	 ground based 	&	 Y 	&	 3.10$^{+0.40}_{-0.40}$ 	&	 81.26$^{+19.20}_{-19.20}$ 	&	 34.00$\pm 28.00$ 	&	 4.28$^{+0.20}_{-0.20}$ 	&	 15\\[0.5ex]
		 HD 182255 	&	 ground based 	&	 Y 	&	 3.90$^{+0.60}_{-0.60}$ 	&	 77.85$^{+22.82}_{-22.82}$ 	&	 39.00$\pm 31.00$ 	&	 4.23$^{+0.20}_{-0.20}$ 	&	 10\\[0.5ex]
		 HD 208057 	&	 ground based 	&	 N 	&	 5.10$^{+0.80}_{-0.80}$ 	&	 72.72$^{+26.96}_{-26.96}$ 	&	 46.00$\pm 33.00$ 	&	 4.15$^{+0.20}_{-0.20}$ 	&	 10\\[0.5ex]
		 HD 215573 	&	 ground based 	&	 N 	&	 4.00$^{+0.70}_{-0.70}$ 	&	 62.78$^{+27.08}_{-27.08}$ 	&	 58.00$\pm 28.00$ 	&	 4.09$^{+0.20}_{-0.20}$ 	&	 10\\[0.5ex]
 \enddata
 \tablecomments{The type of data used for the asteroseismic analysis is provided in the second column, the third column gives the mass of the star, while the fourth column indicates if the star is a known/suspected binary (Y) or not (N). Column five to seven lists the three age indicators available or derived from literature values (see Appendix~\ref{App:Age_indicators}), while the last column gives the source of the age indicator and mass.\\
 References: (1) \cite{Pedersen2022}, (2) \cite{Szewczuk2022}, (3) \cite{Wu2020}, (4) \cite{Buysschaert2018}, (5) \cite{Wu2019}, (6) \cite{Zwintz2017}, (7) \cite{Kallinger2017}, (8) \cite{DeCat2002}, (9) \cite{Gebruers2021}, (10) \cite{Hubrig2006}, (11) \cite{Sodor2014}, (12) \cite{Fedurco2020}, (13) \cite{Aerts2005}, (14) \cite{Briquet2007}, (15) \cite{DeCat2007b}.}
\end{deluxetable*}
}

\section{Rotational spin parameters}\label{sec:Spin}

\begin{figure*}
\center
\includegraphics[width=0.75\linewidth]{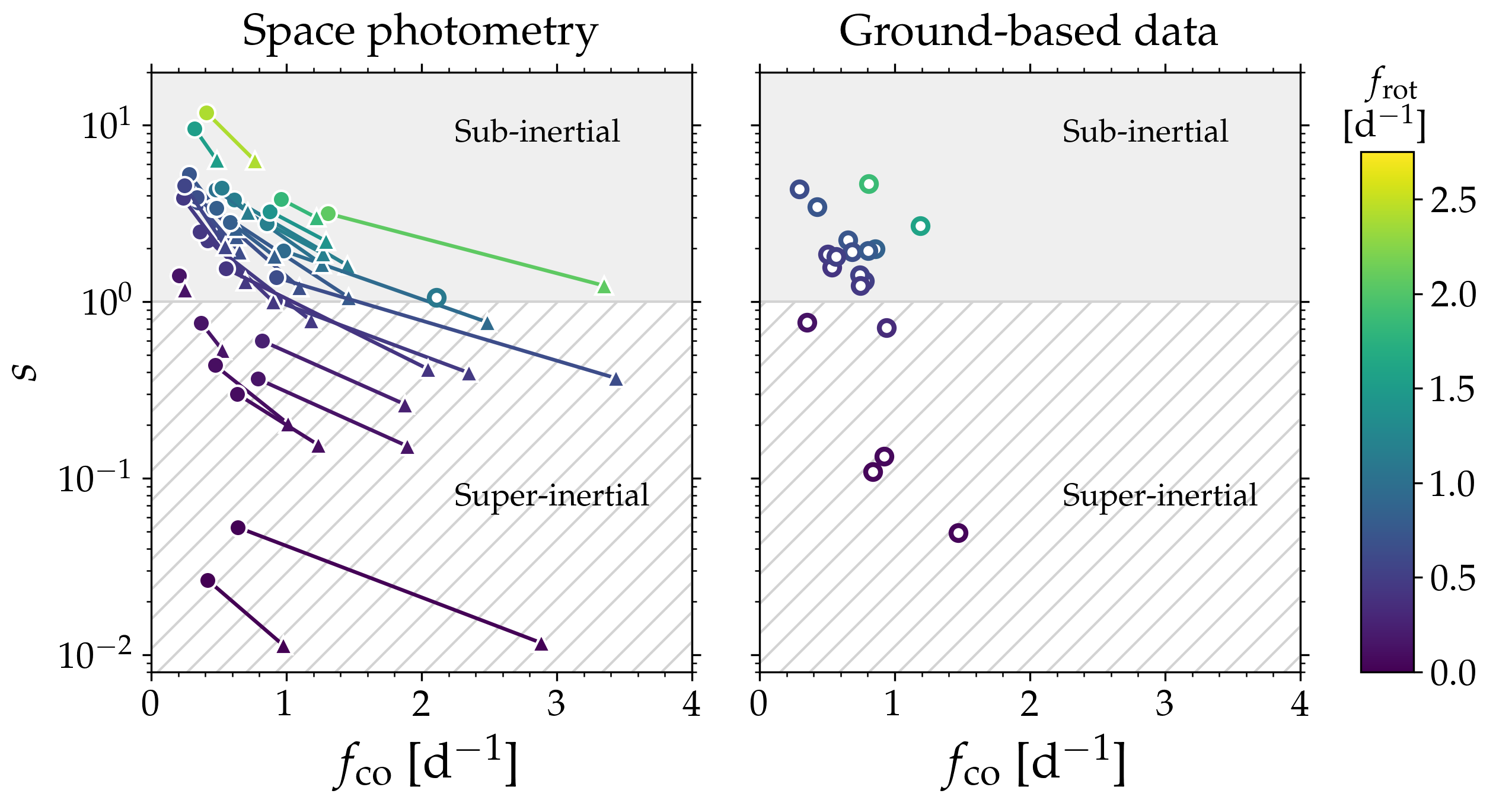}
\caption{Spin parameter as a function of oscillation frequency in the co-rotating frame of reference. Stars for which inferences were made based on space photometry are shown in the left panel, while stars with ground based data are shown on the right. For stars with period spacing patterns, the lowest (circle) and highest (triangle) frequency in the period spacing pattern are connected by lines. Open symbols correspond to stars without period spacing patterns. The colors indicate the derived rotation frequencies of the stars. The super-inertial regime is indicated by the hatched region, while the light-grey shaded region shows part of the sub-inertial regime.}\label{fig:Spin}
\end{figure*}

In the absence of rotation, all oscillation modes with $m = -\ell, \dots, \ell$ and the same radial order $n$ share the same oscillation frequency. For a rotating star, however, this degeneracy is lifted, and a rotational splitting of the oscillation frequency into a $2\ell +1$ multiplet occurs due to the effects of the Coriolis force \citep{Aerts2010}. The faster the rotation, the larger the observed splitting will be until the regularity of the multiplets are no longer clearly visible. Depending on how fast the star is rotating, the contribution from the Coriolis force can be treated as a perturbation to the oscillation equations. However, if the rotation frequency becomes comparable to or larger than the oscillation frequency this is no longer the case, and the g-mode oscillations occur in the sub-inertial instead of the super-inertial regime. In both regimes, the oscillations behave as gravito-inertial modes in radiative regions, but either propagate as inertial modes (sub-inertial regime) or are evanescent (super-inertial regime) in convective zones \citep[e.g.][]{Mathis2014}. To quantify these regimes, we use the spin parameter $s$, which is also the inverse of the Rossby number and describes the relative importance of rotation

\begin{equation}
s = \frac{2 f_{\rm rot}}{f_{\rm co}}.
	\label{Eq:Spin}
\end{equation}

Here $f_{\rm rot}$ is the rotation frequency of the star, and $f_{\rm co}$ is the frequency of the oscillation in the co-rotating frame. It is related to the frequency in the inertial frame of reference $f_{\rm in}$, i.e. the observed oscillation frequency, through the transformation between the two reference frames 

\begin{equation}
f_{\rm in} = f_{\rm co} + m f_{\rm rot}.
	\label{Eq:f_inertial}
\end{equation}

For four of the 52 SPB stars (KIC~8264293, KIC~8322482, HD~48012, and HD~24587), an equatorial rotational velocity $V_{\rm eq}$ was provided instead of $f_{\rm rot}$ from asteroseismology. Common for these four stars is that the $V_{\rm eq}$ values were derived under the assumption of rigid rotation, and we therefore convert them to rotation frequencies using $f_{\rm rot} = V_{\rm eq} / (2 \pi R)$, where $R$ is the radius of the star. For stars without a known radius we calculate it, when possible, using either $R = \sqrt{L/ (4 \pi \sigma_{\rm SB} T_{\rm eff}^4)}$ or $R = \sqrt{GM / g}$ depending on what information is available.

In Fig.~\ref{fig:Spin} we show the spin parameters calculated for the SPB stars with an asteroseismic rotation frequency measured from space photometry (left) and ground-based data (right). For stars with period spacing patterns the highest (triangle) and lowest (circle) oscillation frequency in the pattern is connected by a line to show the range in spin parameters for the pattern. For stars without period spacing patterns, we instead calculate the spin parameter for the highest amplitude oscillation mode (open circles). The colors indicate how fast the star is rotating near the convective core, where the g-modes have their dominant probing power. As seen in the figure, 25 of the 32 stars with period spacing patterns have at least a part of their pattern falling inside the sub-inertial regime. For 15 of the 20 SPB stars without period spacing patterns, the dominant oscillation mode likewise fall within the sub-inertial regime. This is in line with previous results for SPB stars and their lower mass counterpart the $\gamma$ ~Dor stars \citep{VanReeth2016, Aerts2017,Li2020}, which likewise oscillate in g-modes and have period spacing patterns. We list the derived spin parameters shown in Fig.~\ref{fig:Spin} in Table~\ref{Tab:rot_parameters} along with the parameters used to derive them.

\cite{Pedersen2021} took the effects of the Coriolis force on the oscillation frequencies into account when calculating the theoretical oscillation frequencies with the stellar oscillation code \texttt{GYRE} \citep{Townsend2013,Townsend2018,Goldstein2020} and when deriving the observed rotational frequency from the slope of the period spacing patterns by adopting the Traditional Approximation of Rotation \citep[TAR,][]{Eckart1960,Lee1987,Bildsten1996,Townsend2005}. In this approximation, the horizontal component of the rotation vector is ignored, allowing the pulsation equations to be separated into their spherical coordinates $(r, \theta, \phi)$, while also assuming that the star is sufficiently slowly rotating such that the centrifugal force can be ignored \citep{Lee1987,Bildsten1996,Lee1997}. The TAR was previously adopted in the estimation of the rotation frequency and/or the asteroseismic modeling for 46 out of the 52 SPB stars discussed in this work, whereas the rotation was considered small enough for the Coriolis force to be treated as a perturbation to the oscillation equations for six of the stars. Under the assumption of rigid rotation, five of these six stars rotate less than 6\,\% of their Keplerian breakup velocity, in which case 1st order perturbation methods are sufficient for the frequency range of the observed oscillations, see Fig.~5 of \cite{Ballot2010}. For the sixth star, the rotation rate with respect to the breakup velocity could not be estimated because the mass of the star is unknown.

For increasingly faster rotating stars, the TAR will start to break down as the importance of the centrifugal force increases and the star becomes more and more deform. The effects on the deformation of the star are smaller near the convective cores where the gravity and gravito-inertial modes are most sensitive, implying that the TAR may still be adequate even in the cases of rapid rotation for some of the oscillation modes. One study comparing the period spacings obtained with the TAR to those derived when the 2D effects of rotation are accounted for was carried out by \cite{Ouazzani2017}, following a previous similar study by \cite{Ballot2012}. Figure~1 of \cite{Ouazzani2017} shows that the largest deviations between the TAR and the 2D treatment are found for the retrograde ($m < 0$) modes, while the differences are smaller than $\approx 4\%$ and $\approx 0.8\%$ for zonal ($m=0$) and prograde ($m > 0$) modes, respectively, for $f_{\rm rot} < 2\,{\rm d}^{-1}$. For rotation frequencies smaller than 1\,d$^{-1}$, the differences remain smaller than $0.2\%$ for all azimuthal orders. For 40 out of the 52 SPB stars the derived rotation frequencies are smaller than 1\,d$^{-1}$, and only two of the stars have $f_{\rm rot} \approx 2$\,d$^{-1}$. We therefore consider the TAR to be a good assumption for the majority of the 52~SPB stars considered in this work.

{\movetabledown=0.5in
\tabcolsep=5pt
\begin{longrotatetable}
\begin{deluxetable*}{lcccccccccccc}
\tablecaption{Additional asteroseismic information on the SPB star sample.
 \label{Tab:rot_parameters}}
\tablewidth{700pt}
\tabletypesize{\small}
\tablehead{
\colhead{KIC} & 
\colhead{$\ell$} & 
\colhead{$m$} & 
\colhead{$f_{\rm rot}$} &
\colhead{$f_{\rm in}^{\rm 1}$} & 
\colhead{$f_{\rm in}^{\rm 2}$} &
 \colhead{$s^{\rm 1}$} &
  \colhead{$s^{\rm 2}$} &
   \colhead{$R$} &
   \colhead{$V_{\rm eq}\sin i$} &
\colhead{$i$} &
\colhead{References}\\
& & & \colhead{(d$^{-1}$)} & \colhead{(d$^{-1}$)} & \colhead{(d$^{-1}$)} & & & \colhead{($R_\odot$)} & \colhead{(km s$^{-1}$)}
& \colhead{($^{\rm o}$)}
} 
\startdata
		 KIC 3240411	&	 1	&	 0 	&	 1.166$^{+0.008}_{-0.008}$ 	&	 0.61554$\pm$0.00003 	&	 1.45306$\pm$0.00002 	&	 3.79$^{+0.03}_{-0.03}$ 	&	 1.60$^{+0.01}_{-0.01}$ 	&	 3.31$^{+0.18}_{-0.18}$ 	&	 22 $\pm$ 5 	&	 6.3$^{+1.5}_{-1.5}$ 	&	 1,2,3\\[0.5ex]
		 KIC 3459297	&	 1	&	 1 	&	 0.623$^{+0.025}_{-0.025}$ 	&	 0.89350$\pm$0.00002 	&	 1.12832$\pm$0.00003 	&	 4.61$^{+1.25}_{-1.25}$ 	&	 2.47$^{+0.42}_{-0.42}$ 	&	 4.00$^{+0.29}_{-0.29}$ 	&	 106 $\pm$ 14 	&	 55.9$^{+11.8}_{-11.8}$ 	&	 1,2,4\\[0.5ex]
		 KIC 3865742	&	 1	&	 1 	&	 0.625$^{+0.022}_{-0.022}$ 	&	 0.97001$\pm$0.00003 	&	 1.27767$\pm$0.00007 	&	 3.63$^{+1.51}_{-1.51}$ 	&	 1.92$^{+0.53}_{-0.53}$ 	&	 5.78$^{+0.83}_{-0.83}$ 	&	 129 $\pm$ 26 	&	 42.3$^{+13.7}_{-13.7}$ 	&	 1,2,5\\[0.5ex]
		 KIC 4930889	&	 1	&	 1 	&	 0.740$^{+0.008}_{-0.008}$ 	&	 1.02072$\pm$0.00007 	&	 1.37295$\pm$0.00006 	&	 5.27$^{+0.16}_{-0.16}$ 	&	 2.34$^{+0.04}_{-0.04}$ 	&	 3.64$^{+0.15}_{-0.15}$ 	&	 116 $\pm$ 6 	&	 58.2$^{+6.6}_{-6.6}$ 	&	 1,2,3\\[0.5ex]
		 KIC 4936089	&	 1	&	 0 	&	 0.144$^{+0.025}_{-0.025}$ 	&	 0.78792$\pm$0.00006 	&	 1.89420$\pm$0.00006 	&	 0.37$^{+0.12}_{-0.12}$ 	&	 0.15$^{+0.05}_{-0.05}$ 	&	 3.92$^{+0.51}_{-0.51}$ 	&	 48 $\pm$ 7 	&	 \dots 	&	1,2,6\\[0.5ex]
		 KIC 4939281	&	 1	&	 1 	&	 0.772$^{+0.001}_{-0.001}$ 	&	 1.22779$\pm$0.00013 	&	 2.23421$\pm$0.00004 	&	 3.38$^{+0.01}_{-0.01}$ 	&	 1.06$^{+0.00}_{-0.00}$ 	&	 5.38$^{+0.38}_{-0.38}$ 	&	 115 $\pm$ 20 	&	 33.1$^{+7.3}_{-7.3}$ 	&	 1,2,4\\[0.5ex]
		 KIC 5309849	&	 1	&	 1 	&	 0.462$^{+0.005}_{-0.005}$ 	&	 0.87635$\pm$0.00002 	&	 1.64622$\pm$0.00001 	&	 2.23$^{+0.04}_{-0.04}$ 	&	 0.78$^{+0.01}_{-0.01}$ 	&	 5.55$^{+0.75}_{-0.75}$ 	&	 \dots 	&	 \dots 	&	 1,2,7\\[0.5ex]
		 KIC 5941844	&	 1	&	 0 	&	 0.952$^{+0.037}_{-0.037}$ 	&	 0.97725$\pm$0.00007 	&	 2.48880$\pm$0.00013 	&	 1.95$^{+0.15}_{-0.15}$ 	&	 0.76$^{+0.06}_{-0.06}$ 	&	 2.25$^{+0.07}_{-0.07}$ 	&	 34 $\pm$ 4 	&	 18.2$^{+3.0}_{-3.0}$ 	&	 1,2,5\\[0.5ex]
		 KIC 6352430	&	 1	&	 1 	&	 0.659$^{+0.062}_{-0.062}$ 	&	 0.99772$\pm$0.00002 	&	 1.75446$\pm$0.00002 	&	 3.90$^{+2.76}_{-2.76}$ 	&	 1.20$^{+0.45}_{-0.45}$ 	&	 2.89$^{+0.48}_{-0.48}$ 	&	 70 $\pm$ 2 	&	 41.9$^{+13.1}_{-13.1}$ 	&	 1,2,8\\[0.5ex]
		 KIC 6462033	&	 1	&	 0 	&	 0.634$^{+0.012}_{-0.012}$ 	&	 0.92520$\pm$0.00001 	&	 3.43849$\pm$0.00002 	&	 1.37$^{+0.12}_{-0.12}$ 	&	 0.37$^{+0.03}_{-0.03}$ 	&	 6.66$^{+0.24}_{-0.24}$ 	&	 75 $\pm$ 19 	&	 20.3$^{+6.1}_{-6.1}$ 	&	 1,2,5\\[0.5ex]
		 KIC 6780397	&	 1	&	 0 	&	 0.465$^{+0.250}_{-0.250}$ 	&	 0.91610$\pm$0.00003 	&	 2.35034$\pm$0.00006 	&	 1.02$^{+0.45}_{-0.45}$ 	&	 0.40$^{+0.18}_{-0.18}$ 	&	 5.56$^{+1.00}_{-1.00}$ 	&	 120 $\pm$ 0 	&	 67.9$^{+14.0}_{-14.0}$ 	&	 1,2,9\\[0.5ex]
		 KIC 7630417	&	 1	&	 1 	&	 0.451$^{+0.000}_{-0.000}$ 	&	 0.81281$\pm$0.00001 	&	 1.35478$\pm$0.00003 	&	 2.49$^{+0.11}_{-0.11}$ 	&	 1.00$^{+0.03}_{-0.03}$ 	&	 6.31$^{+0.08}_{-0.08}$ 	&	 135 $\pm$ 16 	&	 69.4$^{+10.4}_{-10.4}$ 	&	 1,2,6\\[0.5ex]
		 KIC 7760680	&	 1	&	 1 	&	 0.453$^{+0.022}_{-0.022}$ 	&	 0.68838$\pm$0.00005 	&	 1.15034$\pm$0.00008 	&	 3.86$^{+0.71}_{-0.71}$ 	&	 1.30$^{+0.13}_{-0.13}$ 	&	 2.89$^{+0.05}_{-0.05}$ 	&	 62 $\pm$ 5 	&	 68.9$^{+9.9}_{-9.9}$ 	&	 1,2,10\\[0.5ex]
		 KIC 8057661	&	 1	&	 0 	&	 0.426$^{+0.142}_{-0.142}$ 	&	 0.55151$\pm$0.00002 	&	 2.04747$\pm$0.00007 	&	 1.54$^{+0.54}_{-0.54}$ 	&	 0.42$^{+0.15}_{-0.15}$ 	&	 6.57$^{+0.30}_{-0.30}$ 	&	 35 $\pm$ 8 	&	 14.1$^{+9.0}_{-9.0}$ 	&	 1,2,5\\[0.5ex]
		 KIC 8255796	&	 1	&	 1 	&	 0.146$^{+0.006}_{-0.006}$ 	&	 0.35248$\pm$0.00003 	&	 0.39565$\pm$0.00003 	&	 1.41$^{+0.07}_{-0.07}$ 	&	 1.17$^{+0.06}_{-0.06}$ 	&	 5.93$^{+0.47}_{-0.47}$ 	&	 \dots 	&	 \dots 	&	 1,2\\[0.5ex]
		 KIC 8264293	&	 1	&	 1 	&	 2.412$^{+0.171}_{-0.159}$ 	&	 2.82140$\pm$0.00009 	&	 3.17810$\pm$0.00010 	&	 11.77$^{+5.75}_{-5.35}$ 	&	 6.29$^{+1.85}_{-1.72}$ 	&	 2.03$^{+0.06}_{-0.04}$ 	&	 284 $\pm$ 13 	&	 \dots 	&	 11,6\\[0.5ex]
		 KIC 8324482	&	 1	&	 0 	&	 0.017$^{+0.001}_{-0.001}$ 	&	 0.63845$\pm$0.00017 	&	 2.88685$\pm$0.00017 	&	 0.05$^{+0.00}_{-0.00}$ 	&	 0.01$^{+0.00}_{-0.00}$ 	&	 3.06$^{+0.09}_{-0.07}$ 	&	 \dots 	&	 \dots 	&	 12\\[0.5ex]
		 KIC 8381949	&	 1	&	 1 	&	 0.817$^{+0.182}_{-0.182}$ 	&	 1.29847$\pm$0.00004 	&	 1.44268$\pm$0.00003 	&	 3.39$^{+0.21}_{-0.21}$ 	&	 2.61$^{+0.13}_{-0.13}$ 	&	 5.58$^{+0.91}_{-0.91}$ 	&	 245 $\pm$ 21 	&	 90.0$^{+10.3}_{-10.3}$ 	&	 1,2,6\\[0.5ex]
		 KIC 8459899	&	 1	&	 0 	&	 0.140$^{+0.075}_{-0.075}$ 	&	 0.36985$\pm$0.00006 	&	 0.52630$\pm$0.00008 	&	 0.76$^{+0.30}_{-0.30}$ 	&	 0.53$^{+0.21}_{-0.21}$ 	&	 3.88$^{+0.30}_{-0.30}$ 	&	 53 $\pm$ 4 	&	 \dots 	&	 1,2,3\\[0.5ex]
		 KIC 8714886	&	 1	&	 -1 	&	 0.246$^{+0.024}_{-0.024}$ 	&	 0.57215$\pm$0.00002 	&	 1.63250$\pm$0.00003 	&	 0.60$^{+0.06}_{-0.06}$ 	&	 0.26$^{+0.02}_{-0.02}$ 	&	 4.78$^{+0.04}_{-0.04}$ 	&	 52 $\pm$ 9 	&	 61.0$^{+12.6}_{-12.6}$ 	&	 1,2,6\\[0.5ex]
		 KIC 8766405	&	 1	&	 1 	&	 0.563$^{+0.042}_{-0.042}$ 	&	 0.80955$\pm$0.00008 	&	 1.11239$\pm$0.00004 	&	 4.56$^{+1.81}_{-1.81}$ 	&	 2.05$^{+0.47}_{-0.47}$ 	&	 5.59$^{+0.27}_{-0.27}$ 	&	 240 $\pm$ 12 	&	 \dots 	&	 1,2,3\\[0.5ex]
		 KIC 9020774	&	 1	&	 1 	&	 1.030$^{+0.024}_{-0.024}$ 	&	 1.50818$\pm$0.00002 	&	 2.29514$\pm$0.00005 	&	 4.31$^{+1.44}_{-1.44}$ 	&	 1.63$^{+0.30}_{-0.30}$ 	&	 2.60$^{+0.23}_{-0.23}$ 	&	 129 $\pm$ 28 	&	 68.8$^{+14.3}_{-14.3}$ 	&	 1,2,4\\[0.5ex]
		 KIC 9715425	&	 1	&	 1 	&	 0.826$^{+0.190}_{-0.190}$ 	&	 1.40955$\pm$0.00013 	&	 1.73851$\pm$0.00016 	&	 2.83$^{+0.79}_{-0.79}$ 	&	 1.81$^{+0.39}_{-0.39}$ 	&	 4.63$^{+0.42}_{-0.42}$ 	&	 122 $\pm$ 12 	&	 39.4$^{+10.6}_{-10.6}$ 	&	 1,2,6\\[0.5ex]
		 KIC 10526294	&	 1	&	 0 	&	 0.103$^{+0.005}_{-0.057}$ 	&	 0.47222$\pm$0.00006 	&	 1.01341$\pm$0.00004 	&	 0.44$^{+0.02}_{-0.24}$ 	&	 0.20$^{+0.01}_{-0.11}$ 	&	 4.37$^{+1.01}_{-1.23}$ 	&	 18 $\pm$ 4 	&	 52.1$^{+15.2}_{-15.2}$ 	&	 1,2,13\\[0.5ex]
		 KIC 10536147	&	 1	&	 1 	&	 1.150$^{+0.006}_{-0.006}$ 	&	 1.67156$\pm$0.00013 	&	 1.86423$\pm$0.00014 	&	 4.41$^{+1.09}_{-1.09}$ 	&	 3.22$^{+0.62}_{-0.62}$ 	&	 3.70$^{+0.17}_{-0.17}$ 	&	 163 $\pm$ 18 	&	 48.5$^{+10.3}_{-10.3}$ 	&	 1,2,5\\[0.5ex]
		 KIC 11360704	&	 1	&	 1 	&	 1.189$^{+0.119}_{-0.119}$ 	&	 2.04329$\pm$0.00006 	&	 2.46320$\pm$0.00001 	&	 2.78$^{+0.19}_{-0.19}$ 	&	 1.87$^{+0.10}_{-0.10}$ 	&	 4.15$^{+0.33}_{-0.33}$ 	&	 303 $\pm$ 12 	&	 \dots 	&	 1,2,6\\[0.5ex]
		 KIC 11971405	&	 1	&	 1 	&	 1.535$^{+0.002}_{-0.002}$ 	&	 1.85641$\pm$0.00008 	&	 2.01955$\pm$0.00016 	&	 9.55$^{+0.46}_{-0.46}$ 	&	 6.33$^{+0.21}_{-0.21}$ 	&	 3.14$^{+0.01}_{-0.01}$ 	&	 242 $\pm$ 14 	&	 82.9$^{+7.0}_{-7.0}$ 	&	 1,2,4\\[0.5ex]
		 KIC 12258330	&	 1	&	 0 	&	 2.065$^{+0.002}_{-0.002}$ 	&	 1.30679$\pm$0.00009 	&	 3.35279$\pm$0.00018 	&	 3.16$^{+0.08}_{-0.08}$ 	&	 1.23$^{+0.03}_{-0.03}$ 	&	 2.29$^{+0.02}_{-0.02}$ 	&	 130 $\pm$ 8 	&	 32.9$^{+2.5}_{-2.5}$ 	&	 1,2,3\\[0.5ex]
		 HD 43317	&	 1	&	 -1 	&	 1.114$^{+0.000}_{-0.000}$ 	&	 0.99541$\pm$0.00003 	&	 \dots 	&	 1.06$^{+0.00}_{-0.00}$ 	&	 \dots 	&	 3.39$^{+0.08}_{-0.03}$ 	&	 115 $\pm$ 9 	&	 37.0$^{+3.6}_{-3.4}$ 	&	 14,15\\[0.5ex]
		 HD 50230	&	 1	&	 0 	&	 0.096$^{+0.096}_{-0.064}$ 	&	 0.63683$\pm$0.00005 	&	 1.23685$\pm$0.00004 	&	 0.30$^{+0.30}_{-0.20}$ 	&	 0.15$^{+0.15}_{-0.10}$ 	&	 5.68$^{+0.18}_{-0.13}$ 	&	 7 $\pm$ 2 	&	 14.6$^{+12.5}_{-12.1}$ 	&	 16,17,18\\[0.5ex]
		 HD 48012	&	 1	&	 1 	&	 1.830$^{+0.130}_{-0.121}$ 	&	 2.79000$\pm$0.00200 	&	 3.05400$\pm$0.00800 	&	 3.81$^{+0.79}_{-0.73}$ 	&	 2.99$^{+0.53}_{-0.49}$ 	&	 2.79$^{+0.08}_{-0.06}$ 	&	 225 $\pm$ 20 	&	 60.7$^{+9.7}_{-9.7}$ 	&	 19\\[0.5ex]
		 HD 261810	&	 1	&	 1 	&	 1.427$^{+0.101}_{-0.094}$ 	&	 2.30500$\pm$0.00500 	&	 2.72300$\pm$0.00300 	&	 3.25$^{+0.61}_{-0.56}$ 	&	 2.20$^{+0.33}_{-0.31}$ 	&	 3.27$^{+0.10}_{-0.07}$ 	&	 180 $\pm$ 15 	&	 49.7$^{+8.2}_{-7.9}$ 	&	19\\[0.5ex]
		 HD 201433	&	 1	&	 1 	&	 0.006$^{+0.000}_{-0.000}$ 	&	 0.42340$\pm$0.00140 	&	 0.98431$\pm$0.00029 	&	 0.03$^{+0.00}_{-0.00}$ 	&	 0.01$^{+0.00}_{-0.00}$ 	&	 2.77$^{+0.18}_{-0.18}$ 	&	 8 $\pm$ 2 	&	 ($68\pm5$) 	&	 20\\[0.5ex]
		 HD 1976	&	 1	&	 1 	&	 0.644$^{+0.100}_{-0.100}$ 	&	 0.93895$\pm$0.00004 	&	 \dots 	&	 4.36$^{+2.16}_{-2.16}$ 	&	 \dots 	&	 4.00$^{+1.30}_{-1.30}$ 	&	 143 $\pm$ 9 	&	 90.0$^{+0.0}_{-12.6}$ 	&	 21,22\\[0.5ex]
		 HD 21071	&	 1	&	 0 	&	 1.594$^{+0.004}_{-0.004}$ 	&	 1.18843$\pm$0.00001 	&	 \dots 	&	 2.68$^{+0.01}_{-0.01}$ 	&	 \dots 	&	 2.70$^{+0.70}_{-0.70}$ 	&	 51 $\pm$ 4 	&	 13.5$^{+5.5}_{-5.6}$ 	&	 21,22,5\\[0.5ex]
		 HD 24587	&	 1	&	 1 	&	 0.731$^{+0.036}_{-0.036}$ 	&	 1.15690$\pm$0.00060 	&	 \dots 	&	 3.44$^{+0.46}_{-0.46}$ 	&	 \dots 	&	 2.50$^{+0.60}_{-0.60}$ 	&	 28 $\pm$ 4 	&	 17.6$^{+6.8}_{-6.8}$ 	&	 21,23,24\\[0.5ex]
		 HD 25558	&	 1	&	 0 	&	 0.731$^{+0.036}_{-0.036}$ 	&	 0.65265$\pm$0.00002 	&	 \dots 	&	 2.24$^{+0.11}_{-0.11}$ 	&	 \dots 	&	 2.79$^{+0.97}_{-0.97}$ 	&	 28 $\pm$ 2 	&	 15.7$^{+8.8}_{-8.8}$ 	&	 21,22\\[0.5ex]
		 HD 26326	&	 1	&	 0 	&	 0.419$^{+0.054}_{-0.054}$ 	&	 0.53380$\pm$0.00080 	&	 \dots 	&	 1.57$^{+0.20}_{-0.20}$ 	&	 \dots 	&	 3.10$^{+1.00}_{-1.00}$ 	&	 11 $\pm$ 6 	&	 9.6$^{+8.5}_{-8.5}$ 	&	 21,23,24\\[0.5ex]
		 HD 36999	&	 1	&	 -1 	&	 0.036$^{+0.003}_{-0.002}$ 	&	 1.43227$\pm$0.00002 	&	 \dots 	&	 0.05$^{+0.00}_{-0.00}$ 	&	 \dots 	&	 2.24$^{+0.07}_{-0.05}$ 	&	 \dots 	&	 \dots 	&	 25\\[0.5ex]
		 HD 53921	&	 2	&	 -1 	&	 0.334$^{+0.213}_{-0.213}$ 	&	 0.60540$\pm$0.00060 	&	 \dots 	&	 0.71$^{+0.29}_{-0.29}$ 	&	 \dots 	&	 2.60$^{+0.70}_{-0.70}$ 	&	 17 $\pm$ 10 	&	 22.7$^{+15.9}_{-15.7}$ 	&	 21,23,24\\[0.5ex]
		 HD 66181	&	 1	&	 1 	&	 0.046$^{+0.001}_{-0.001}$ 	&	 0.88522$\pm$0.00002 	&	 \dots 	&	 0.11$^{+0.00}_{-0.00}$ 	&	 \dots 	&	 4.32$^{+0.13}_{-0.10}$ 	&	 \dots 	&	 \dots 	&	 25\\[0.5ex]
		 HD 74195	&	 1	&	 0 	&	 0.134$^{+0.036}_{-0.036}$ 	&	 0.35033$\pm$0.00009 	&	 \dots 	&	 0.77$^{+0.20}_{-0.20}$ 	&	 \dots 	&	 4.30$^{+1.40}_{-1.40}$ 	&	 9 $\pm$ 5 	&	 18.0$^{+14.0}_{-13.9}$ 	&	 21,23,24\\[0.5ex]
		 HD 85953	&	 1	&	 -1 	&	 0.511$^{+0.003}_{-0.003}$ 	&	 0.26630$\pm$0.00060 	&	 \dots 	&	 1.32$^{+0.00}_{-0.00}$ 	&	 \dots 	&	 4.90$^{+1.70}_{-1.70}$ 	&	 18 $\pm$ 10 	&	 8.2$^{+7.6}_{-7.5}$ 	&	 21,23,24\\[0.5ex]
		 HD 92287	&	 1	&	 -1 	&	 0.525$^{+0.025}_{-0.025}$ 	&	 0.21480$\pm$0.00007 	&	 \dots 	&	 1.42$^{+0.02}_{-0.02}$ 	&	 \dots 	&	 3.90$^{+1.30}_{-1.30}$ 	&	 41 $\pm$ 16 	&	 23.3$^{+14.0}_{-13.6}$ 	&	 21,23,24\\[0.5ex]
		 HD 97895	&	 1	&	 0 	&	 0.062$^{+0.000}_{-0.000}$ 	&	 0.92263$\pm$0.00001 	&	 \dots 	&	 0.13$^{+0.00}_{-0.00}$ 	&	 \dots 	&	 3.34$^{+0.10}_{-0.07}$ 	&	 \dots 	&	 \dots 	&	 25\\[0.5ex]
		 HD 121190	&	 1	&	 1 	&	 1.876$^{+0.152}_{-0.152}$ 	&	 2.68310$\pm$0.00040 	&	 \dots 	&	 4.65$^{+1.26}_{-1.26}$ 	&	 \dots 	&	 1.70$^{+0.30}_{-0.30}$ 	&	 118 $\pm$ 3 	&	 47.0$^{+11.0}_{-11.0}$ 	&	 21,26,27\\[0.5ex]
		 HD 123515	&	 1	&	 0 	&	 0.654$^{+0.079}_{-0.079}$ 	&	 0.68528$\pm$0.00010 	&	 \dots 	&	 1.91$^{+0.23}_{-0.23}$ 	&	 \dots 	&	 2.20$^{+0.50}_{-0.50}$ 	&	 6 $\pm$ 3 	&	 4.7$^{+3.1}_{-3.2}$ 	&	 21,23,24\\[0.5ex]
		 HD 160762	&	 1	&	 -1 	&	 0.457$^{+0.016}_{-0.016}$ 	&	 0.28671$\pm$0.00027 	&	 \dots 	&	 1.23$^{+0.02}_{-0.02}$ 	&	 \dots 	&	 5.06$^{+0.50}_{-0.50}$ 	&	 6 $\pm$ 1 	&	 2.9$^{+0.6}_{-0.6}$ 	&	 28\\[0.5ex]
		 HD 179588	&	 1	&	 0 	&	 0.855$^{+0.207}_{-0.207}$ 	&	 0.85654$\pm$0.00081 	&	 \dots 	&	 2.00$^{+0.48}_{-0.48}$ 	&	 \dots 	&	 2.20$^{+0.50}_{-0.50}$ 	&	 52 $\pm$ 14 	&	 33.1$^{+14.4}_{-14.3}$ 	&	21,22\\[0.5ex]
		 HD 182255	&	 1	&	 1 	&	 0.468$^{+0.076}_{-0.076}$ 	&	 0.97185$\pm$0.00003 	&	 \dots 	&	 1.86$^{+0.58}_{-0.58}$ 	&	 \dots 	&	 2.60$^{+0.70}_{-0.70}$ 	&	 10 $\pm$ 6 	&	 9.3$^{+8.1}_{-8.1}$ 	&	21,23,24 \\[0.5ex]
		 HD 208057	&	 1	&	 0 	&	 0.779$^{+0.096}_{-0.096}$ 	&	 0.80213$\pm$0.00010 	&	 \dots 	&	 1.94$^{+0.24}_{-0.24}$ 	&	 \dots 	&	 3.20$^{+1.00}_{-1.00}$ 	&	 104 $\pm$ 6 	&	 55.5$^{+13.3}_{-13.2}$ 	&	 21,22,24\\[0.5ex]
		 HD 215573	&	 1	&	 0 	&	 0.508$^{+0.054}_{-0.054}$ 	&	 0.56540$\pm$0.00060 	&	 \dots 	&	 1.80$^{+0.19}_{-0.19}$ 	&	 \dots 	&	 3.00$^{+1.00}_{-1.00}$ 	&	 5 $\pm$ 2 	&	 3.7$^{+3.7}_{-3.9}$ 	&	 21,23,24\\[0.5ex]
		 \enddata
		  \tablecomments{List of mode identifications ($\ell, m$), rotational frequencies $f_{\rm rot}$, and spin parameters $s$ derived from the minimum ($f_{\rm in}^{\rm 1}$, $s^{\rm 1}$) and maximum ($f_{\rm in}^{\rm 2}$, $s^{\rm 2}$) frequency of the period spacing patterns in the in inertial frame of reference. For stars without period spacing patterns, the frequency and spin parameter of the dominant oscillation mode is provided instead. Column 9-11 list the stellar radii, the projected equatorial velocity $V_{\rm eq} \sin i$, and the derived inclination angles achieved under the assumption of rigid rotation, respectively. The last column lists the source of the mode identification, rotation frequency, oscillation frequencies, radii, and $V_{\rm eq} \sin i$ values. Spin parameters and inclination angles were derived in this work. Inclination angles taken from the literature are listed in parenthesis.\\
 References: (1) \cite{Pedersen2021}, (2) \cite{Pedersen2022}, (3) \cite{Lehmann2011}, (4) \cite{Papics2017}, (5) \cite{Gebruers2021}, (6) \cite{Hanes2019}, (7) \cite{Zhang2018}, (8) \cite{Papics2013}, (9) \cite{Frasca2016}, (10) \cite{Papics2015}, (11) \cite{Szewczuk2022}, (12) \cite{Wu2020}, (13) \cite{Papics2014}, (14) \cite{Buysschaert2017}, (15) \cite{Buysschaert2018}, (16) \cite{Degroote2010}, (17) \cite{Degroote2012}, (18) \cite{Wu2019}, (19) \cite{Zwintz2017}, (20) \cite{Kallinger2017}, (21) \cite{Szewczuk2015a}, (22) \cite{DeCat2007}, (23) \cite{DeCat2002}, (24) \cite{Hubrig2006}, (25) \cite{Fedurco2020}, (26) \cite{North1994}, (27) \cite{Aerts2005}, (28) \cite{DeCat2002b}}
\end{deluxetable*}
\end{longrotatetable}
}
\clearpage

\section{Age dependence of rotation}\label{Sec:Rot_vs_Age}

\begin{figure*}
\includegraphics[width=0.5\linewidth]{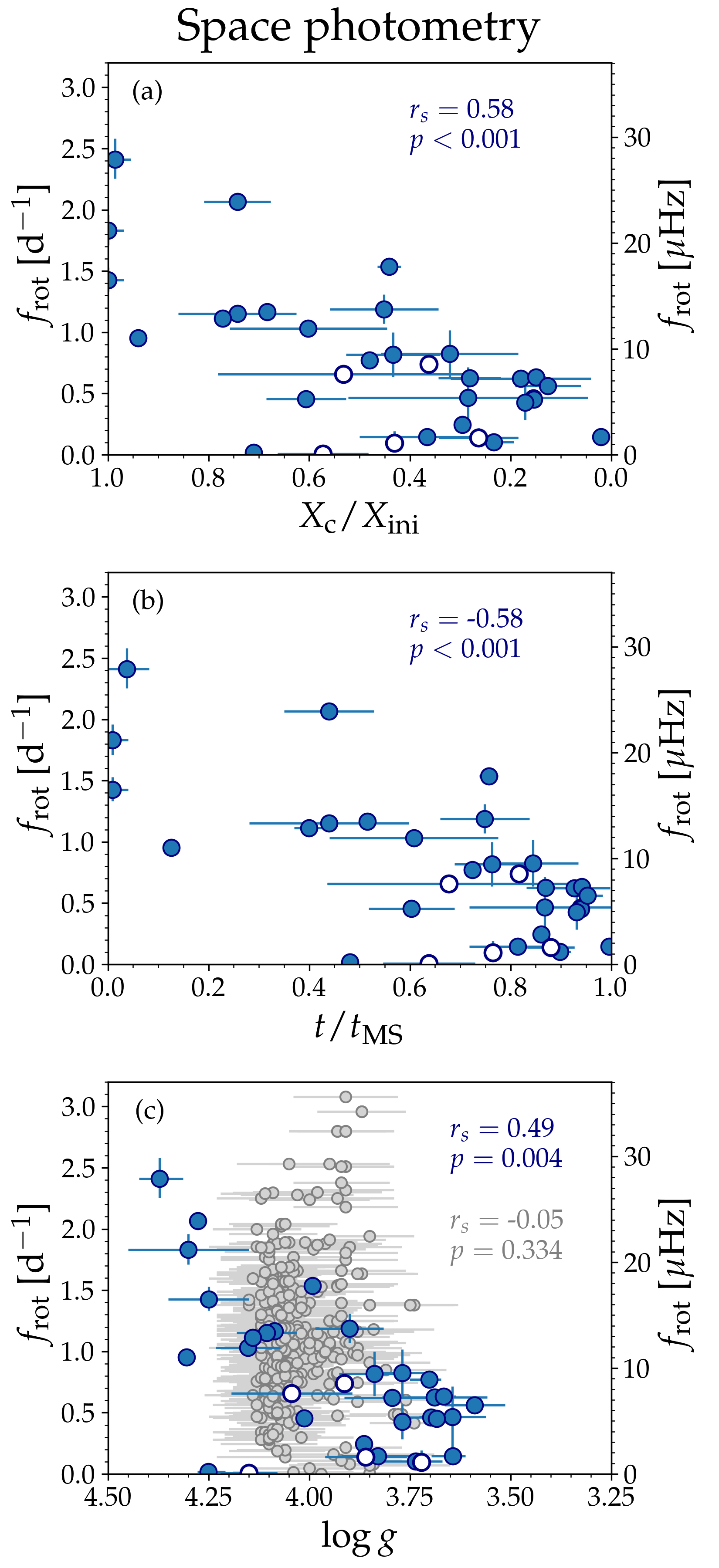}
\includegraphics[width=0.5\linewidth]{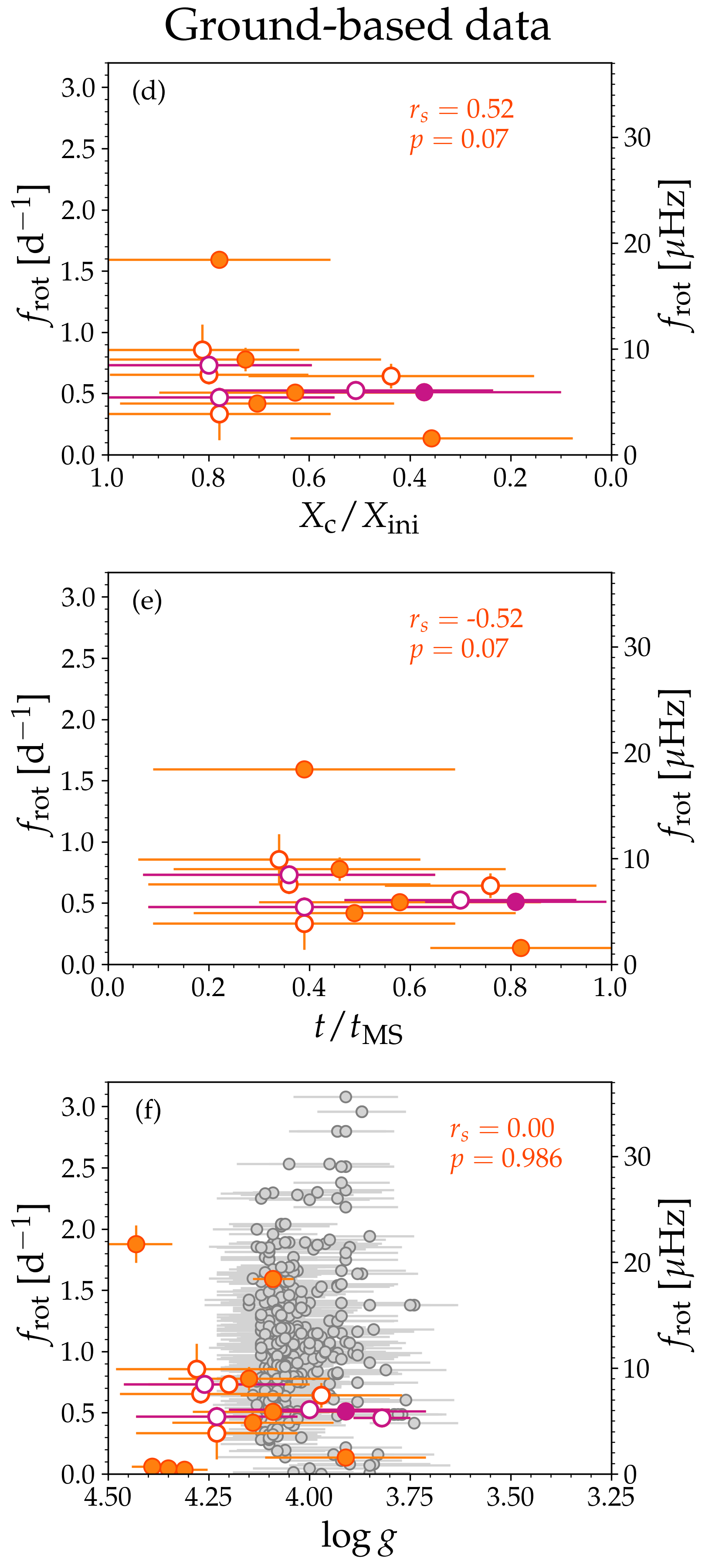}
\caption{Core rotational frequency as a function of main-sequence age (top and middle panels) and surface gravity (bottom panels). The panels on the left-hand side show the results for the 33 SPB stars with space photometry, while the stars with ground based data are shown on the right. Open symbols indicate known or suspected binaries. The four SPB stars with ground-based data for which the inclination angle is found to be inconsistent with the mode identification as discussed in Sect.~\ref{Sec:Inclination} are marked in purple.
The results are compared to the core rotation frequencies of $\approx 300$ $\gamma$~Dor stars derived by \cite{Li2020} in grey. The corresponding Spearman's rank correlation coefficients $r_s$ and their $p$-values are indicated in the subplots.}
	\label{fig:Rot_vs_Age_SPB}
\end{figure*}

\noindent \cite{Aerts2019b} visualized the evolution of the core and surface rotation derived from asteroseismology by collecting the core rotation rates for 1210 stars in the literature, out of which 45 also had measured surface rotation rates. These stars cover masses between 0.72-7.9\,M$_\odot$ and various stages of stellar evolution from main-sequence stars to white dwarfs, but majorly consists of stars on the red giant branch. In their Fig.~4, \cite{Aerts2019b} show the variation of the core rotation as a function of the asteroseismic $\log g$, which is used as a proxy for the age of the stars. Since the generation of this figure, the previously limited sample of main-sequence stars has been significantly increased with the derivation of core rotation rates of more than 600 $\gamma$~Dor stars \citep{Li2020}, while the number of B-stars has remained small. An updated version of Fig.~4 in \cite{Aerts2019b} including this new sample of $\gamma$~ Dor stars is available in \cite{Aerts2021} (Fig.~6), resulting in core-to-surface rotation rates now being available for 111 stars. Here we take a closer look at the core rotation frequency of the 52~SPB stars considered in this work and study its dependence on the age of the star. We note that by core rotation, we mean in fact the rotation just outside of the convective core where the rotational kernel for the g-modes is largest, see bottom panels of Fig.~\ref{fig:gmode_vs_pmode}.

Figure~\ref{fig:Rot_vs_Age_SPB} shows the evolution of the near-core rotation of the SPB stars with space photometry (left) and ground-based data (right) as a function of main-sequence age measured as the ratio of the current to the initial core hydrogen mass fraction of the star $X_{\rm c}/X_{\rm ini}$ (panels (a) and (d)), and as a fraction of the current age to the main-sequence lifetime $t/t_{\rm MS}$ (panels (b) and (e)). Panels (c) and (f) show the corresponding evolution as a function of the surface gravity\footnote{When available, we adopt the asteroseismic $\log g$ instead of the spectroscopic one.} $\log g$, and the near-core rotation is compared to those of the $\approx 300$ $\gamma$~Dor stars (grey) from \cite{Li2020} for which spectroscopic $\log g$ values were available from \cite{Frasca2016}. As seen in Fig.~\ref{fig:Rot_vs_Age_SPB}, the errors on the age indicators for the sample of SPB stars for which the rotation frequency was estimated using data from ground-based telescopes are generally much larger than for the stars with space photometry. This is mainly a result of the lack of an asteroseismic modelling of the stars, in which case a much more precise estimate of the age can be obtained.

For the SPB~stars with space photometry we see a trend that the core rotation decreases as a function of age, independently of whether $X_{\rm c}/X_{\rm ini}$, $t/t_{\rm MS}$, or $\log g$ is used as an age indicator. A similar trend is seen for the sample of 19 SPB stars with asteroseismology based on ground-based data. To quantify this, we calculate the Spearman's rank correlation coefficient $r_s$, which takes values between -1 and 1, corresponding to a strong negative or positive correlation, respectively. 

As indicated in panels (a) and (b) of Fig.~\ref{fig:Rot_vs_Age_SPB}, the Spearman's rank correlation coefficient shows a positive correlation between $f_{\rm rot}$ and both $X_{\rm c}/X_{\rm ini}$ ($r_s=0.58$, $p < 0.001$) and $t/t_{\rm MS}$ ($r_s=-0.58$, $p < 0.001$) for the SPB stars with space photometry, confirming that $f_{\rm rot}$ does indeed decrease with main-sequence age. A similar trend is seen for the SPB stars with ground-based data, but the higher p-values ($p = 0.07$) indicate that there is only weak evidence for rejecting the null-hypothesis that $f_{\rm rot}$ is uncorrelated with $X_{\rm c}/X_{\rm ini}$ and $t/t_{\rm MS}$, compared to the very strong evidence found for the 33 SPB stars with space photometry. Strong evidence is found for a correlation between $f_{\rm rot}$ and $\log g$ for the SPB stars with space photometry ($r_s=0.49$, $p = 0.04$), whereas no such correlation is found for the 19 SPB stars with ground-based data.

One important difference between the top two panels on the right-hand side of Fig.~\ref{fig:Rot_vs_Age_SPB} and panel (f) is that $\log g$ is the only age indicator available for six of the SPB stars with ground-based data. The three slowest rotating stars are among these six stars, and all three of them had their rotation period determined based on data collected from five different ground-based photometric surveys of varying observing cadence not specifically designed for asteroseismology \citep{Fedurco2020}. Nevertheless, while the correlation between the near-core rotation frequency and the three age indicators in panels (a) to (d) have different levels of significance, the overall lack of stars in the upper right corners of these panels show that there are currently no known old main-sequence SPB stars with a high near-core rotation frequency. 

\begin{figure}
\includegraphics[width=\linewidth]{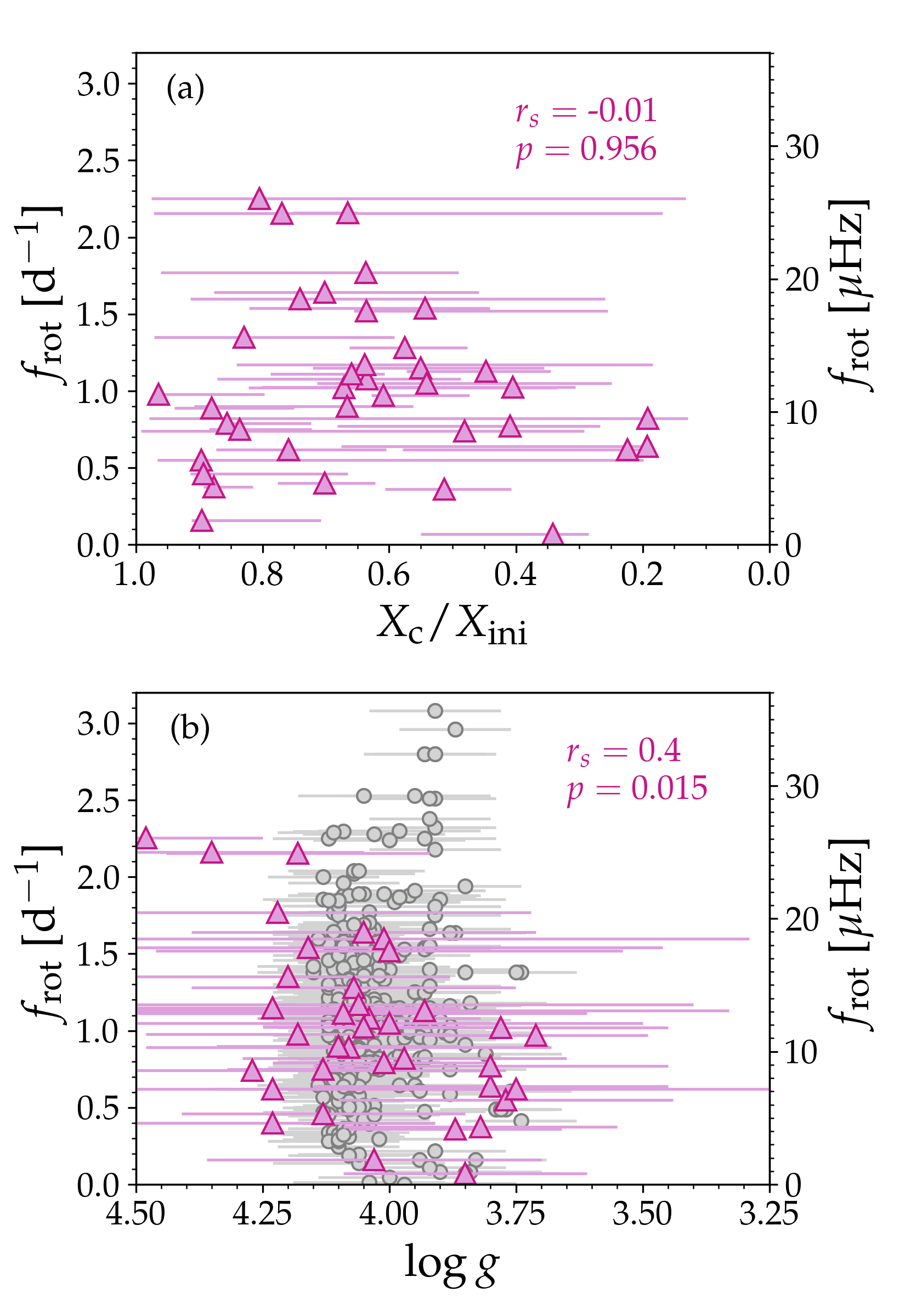}
\caption{Core rotational frequency as a function of main-sequence age (top) and surface gravity (bottom) for 34~$\gamma$~Dor stars (pink), which have been modeled asteroseismically. The $\approx 300$ $\gamma$~Dor stars from \cite{Li2020} are once again shown in grey, and the  Spearman's rank correlation coefficients $r_s$ and their $p$-values are indicated in the subplots.}
	\label{fig:Rot_vs_Age_gDor}
\end{figure}

For the $\gamma$~Dor sample shown in grey in panels (c) and (f), the high $p$-value means that we cannot reject the null-hypothesis that $r_s = 0$, i.e. that $f_{\rm rot}$ and $\log g$ are uncorrelated.
In an attempt to see if this picture changes if we make use of stellar models derived from asteroseismology for $\gamma$~Dor stars, we carry out a similar comparison for 34 $\gamma$~Dor stars in Fig.~\ref{fig:Rot_vs_Age_gDor}, which were recently modeled asterosesimically by \cite{Mombarg2021}. Their core rotation frequencies were taken from \cite{VanReeth2016} and their spectroscopic $\log g$ values from \cite{VanReeth2015}. For this smaller sample of $\gamma$~Dor stars, no correlation is found between $f_{\rm rot}$ and $X_{\rm c}/X_{\rm ini}$, which is consistent with the results for the larger sample from \cite{Li2020}. However, the Spearman's rank correlation coefficient does indicate a positive correlation between $f_{\rm rot}$ and $\log g$ for the 34 modeled $\gamma$~Dor stars, with the associated $p < 0.05$ value indicating that this correlation is significant at the 95\% confidence level. 


\section{Angular momentum transport}\label{Sec:AM}

In order to investigate if the decrease in the near-core rotation frequencies of the sample of SPB stars as a whole as a function of time is evidence for efficient angular momentum transport, we consider the two extreme cases of 1) no angular momentum transport between adjacent cells in the stellar model, and 2) the angular momentum transport occurs instantaneously and enforces rigid rotation throughout the main-sequence evolution. In both cases we assume that the stars start out as rigidly rotating on the zero-age main-sequence.

In the case of no angular momentum transport, we assume that the specific angular momentum 
$j_1(r) = \Omega_1 \left(r\right) i_1 \left(r\right)$ at a time $t_1$ and later time $t_2$ is the same

\begin{equation}
j_1 = j_2.
\end{equation}

Here $r$ is the central radius of the spherical cell, and $\Omega (r)$ and $i(r)$ are the angular rotation frequency and specific moment of inertia of the star at $r$, respectively. We thus obtain the rotation profile $\Omega_2 \left(r \right)$ at time $t_2$ using

\begin{equation}
\Omega_2 \left(r\right)  = \Omega_1 \left(r\right) \frac{i_1 \left(r \right)}{i_2 \left(r \right)}.
\end{equation}

For the second case of instantaneous angular momentum transport, we assume that the total angular momentum $J$ is conserved between different time steps $t_1$ and $t_2$. 

\begin{equation}
J_1 = J_2.
	\label{Eq:AM}
\end{equation}

The total AM is calculated as

\begin{equation}
J = \frac{2}{3} \int_0^{R} r^2 \Omega (r) {\rm d}m,
	\label{Eq:AM}
\end{equation}

\noindent where ${\rm d}m$ is the enclosed mass of a cell. If we assume that the stars remain rigidly rotating throughout their main-sequence evolution, it simply follows that

\begin{equation}
\Omega_2 = \Omega_1 \frac{I_1}{I_2},
\end{equation}

where $I_1$ and $I_2$ are the total moment of inertia at time $t_1$ and $t_2$, respectively.

\begin{figure*}
\center
\includegraphics[width=0.8\linewidth]{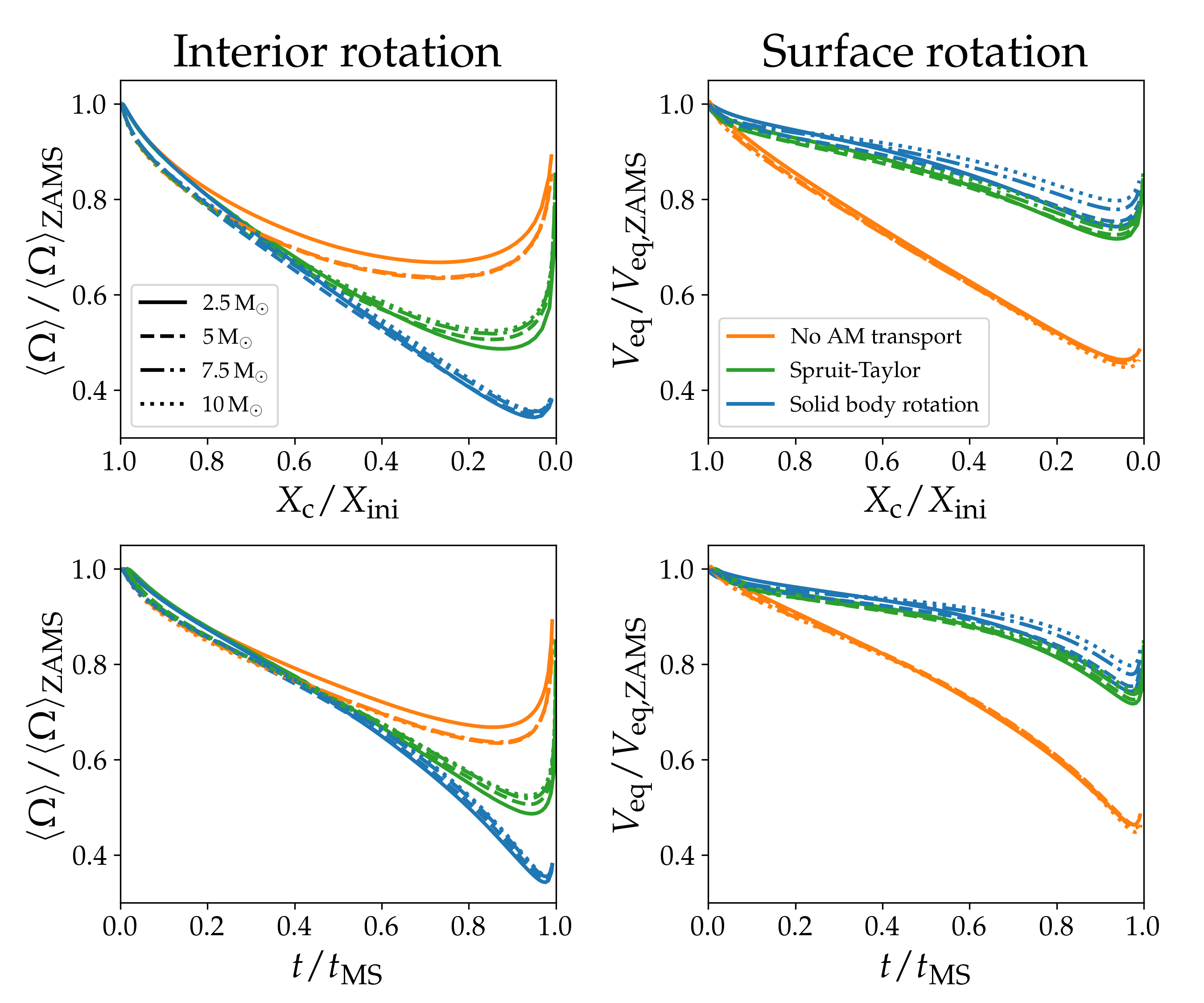}
\caption{Evolution of the average rotation frequencies experienced by the g-modes (left) and the surface equatorial velocities (right) as a function of the two age indicators $X_{\rm c}/X_{\rm ini}$ (top) and $t/t_{\rm MS}$ (bottom). The evolution of the rotation rates are given with respect to the initial rotation rate at the zero-age main-sequence. Different line styles correspond to different stellar masses: 2.5\,M$_\odot$ (full), 5\,M$_\odot$ (dashed), 7.5\,M$_\odot$ (dash-dotted), and 10\,M$_\odot$ (dotted). Results for rigid rotation and no angular momentum transport are shown in blue and orange, respectively. Models with angular momentum transport from the Spruit-Taylor dynamo assuming an initial rotation rate of $10\,\%\,\Omega_{\rm crit}$ are shown in green for comparison.}
	\label{fig:frot_cor_surf}
\end{figure*}

As shown in the bottom panels of Fig.~\ref{fig:gmode_vs_pmode}, the g-modes mainly probe the near core rotation profile of the star and do not provide information about the rotation frequency of the convective core itself. Following \cite{Ouazzani2019}, we therefore calculate the average rotational frequency experienced by the g-modes over their propagation cavity weighted by the Brunt-Vaisala frequency $N$ as

\begin{equation}
\langle \Omega \rangle = \frac{\int_{\rm gc} \Omega \left(r\right) N(r) \frac{{\rm d}r}{r}}{\int_{\rm gc} N(r) \frac{{\rm d}r}{r}},
\end{equation}

and will use this average for comparison with the observed g-mode rotation frequencies rather than a derived central rotation frequency of the convective core. For comparison, we also calculate the corresponding surface equatorial velocities

\begin{equation}
V_{\rm eq} = \Omega_{\rm surf} R,
	\label{Eq:Veq}
\end{equation}

where $\Omega_{\rm surf} = \Omega(R)$.

\subsection{Evolution of the near-core and surface rotation}

As a first step, we derive the evolution of the near-core and surface rotation rates for four different stellar masses 2.5, 5, 7.5, and 10\,M$_\odot$ covering the SPB instability strip. The models were computed with the stellar structure and evolution code \texttt{MESA} version r12115 \citep{Paxton2011,Paxton2013,Paxton2015,Paxton2018,Paxton2019} following the same setup as \cite{Pedersen2021} but excluding winds to ensure conservation of angular momentum, see also Appendix~\ref{App:Age_indicators} for more information. The results of these calculations are shown in Fig.~\ref{fig:frot_cor_surf}. The different line styles corresponds to different stellar masses, while the evolution of the rotation rates assuming rigid rotation and no angular momentum transport are shown in blue and orange, respectively. For comparison, we also show in green what the evolution of the near-core and surface rotation is expected to be for stellar models assuming angular momentum transport from the Spruit-Taylor dynamo \citep{Spruit2002,Paxton2013} assuming an initial rotation rate of $10\%\,\Omega_{\rm crit}$. The corresponding \texttt{MESA} setup is discussed in Appendix~\ref{App:MESA_ST}. Increasing the initial rotation rate causes the evolution of the average rotation frequency and surface velocity assuming angular momentum transport from the Spruit-Taylor dynamo to approach the case of solid body rotation, as illustrated in Fig.~\ref{fig:TS_vary_OmegaCrit} in Appendix~\ref{App:MESA_ST}.

The left-hand side of Fig.~\ref{fig:frot_cor_surf} demonstrates that irrespective of whether the SPB stars experience no or very efficient angular momentum transport, the average rotation frequency decreases with age except at the final stages of the evolution towards hydrogen depletion. The rate of decrease stays nearly the same for the two cases for the first half on the main-sequence evolution until $X_{\rm c}/X_{\rm ini} \approx 0.6$ and $t/t_{\rm MS} \approx 0.4$, and then starts to slow down for the case of no angular momentum transport for the remainder of the evolution. In other words, older SPB stars are expected to have smaller average interior rotation frequencies measured by the g-mode oscillations if the stars are rigid rotators, compared to if no angular momentum transport is happening during the main-sequence evolution. The smallest average rotation frequency $\langle \Omega \rangle / \langle \Omega \rangle_{\rm ZAMS} \approx 0.35$ is obtained at $X_{\rm c}/X_{\rm ini} \approx 0.05$  and $t/t_{\rm MS} \approx 0.098$ in the case of rigid rotation, compared to $\langle \Omega \rangle / \langle \Omega \rangle_{\rm ZAMS} \approx 0.64$ at $X_{\rm c}/X_{\rm ini} \approx 0.27$ and $t/t_{\rm MS} \approx 0.87$ for the case of no angular momentum transport.

The evolution of the equatorial velocities shown on the right-hand side of Fig.~\ref{fig:frot_cor_surf} display much larger differences between the two cases of no and highly efficient angular momentum transport with the surface rotation slowing down much faster for the first case. For the solid body rotation scenario, the equatorial velocity remains nearly constant for the first half of the main-sequence evolution and then reaches a minimum $V_{\rm eq}/V_{\rm eq, ZAMS} \approx 0.77$ at  $X_{\rm c}/X_{\rm ini} \approx 0.06$  and $t/t_{\rm MS} \approx 0.97$. The age at the which the minimum is achieved is the same for the computations without angular momentum transport but the surface velocities are lower ($V_{\rm eq}/V_{\rm eq, ZAMS} \approx 0.46$). We note that the evolution of the rotation rates is only weakly dependent on the stellar mass for the SPB stars, particularly so for $M \geq 5\,$M$_\odot$. The largest mass dependence is seen for the surface rotational velocities assuming rigid rotation.

\subsection{Comparison to the SPB star sample}

To study angular momentum transport on the main-sequence, ideally one would know the starting rotational velocity of the stars on the ZAMS. In their analysis of the evolution of the near-core rotation frequencies of 37 $\gamma$~Dor stars, \cite{Ouazzani2019} accomplished this by combining rotational velocity distributions of three young stellar clusters with disc locking models for the interaction between pre-main-sequence stars and their accretion disks. Here we instead start by comparing the average rotational frequencies of the four youngest SPB stars with space photometry with the rotation distribution of the remaining 29 stars. For this first part we exclude the sample of SPB stars with ground-based data because of their smaller sample size and larger errors on their main-sequence ages.

\begin{figure}
\includegraphics[width=\linewidth]{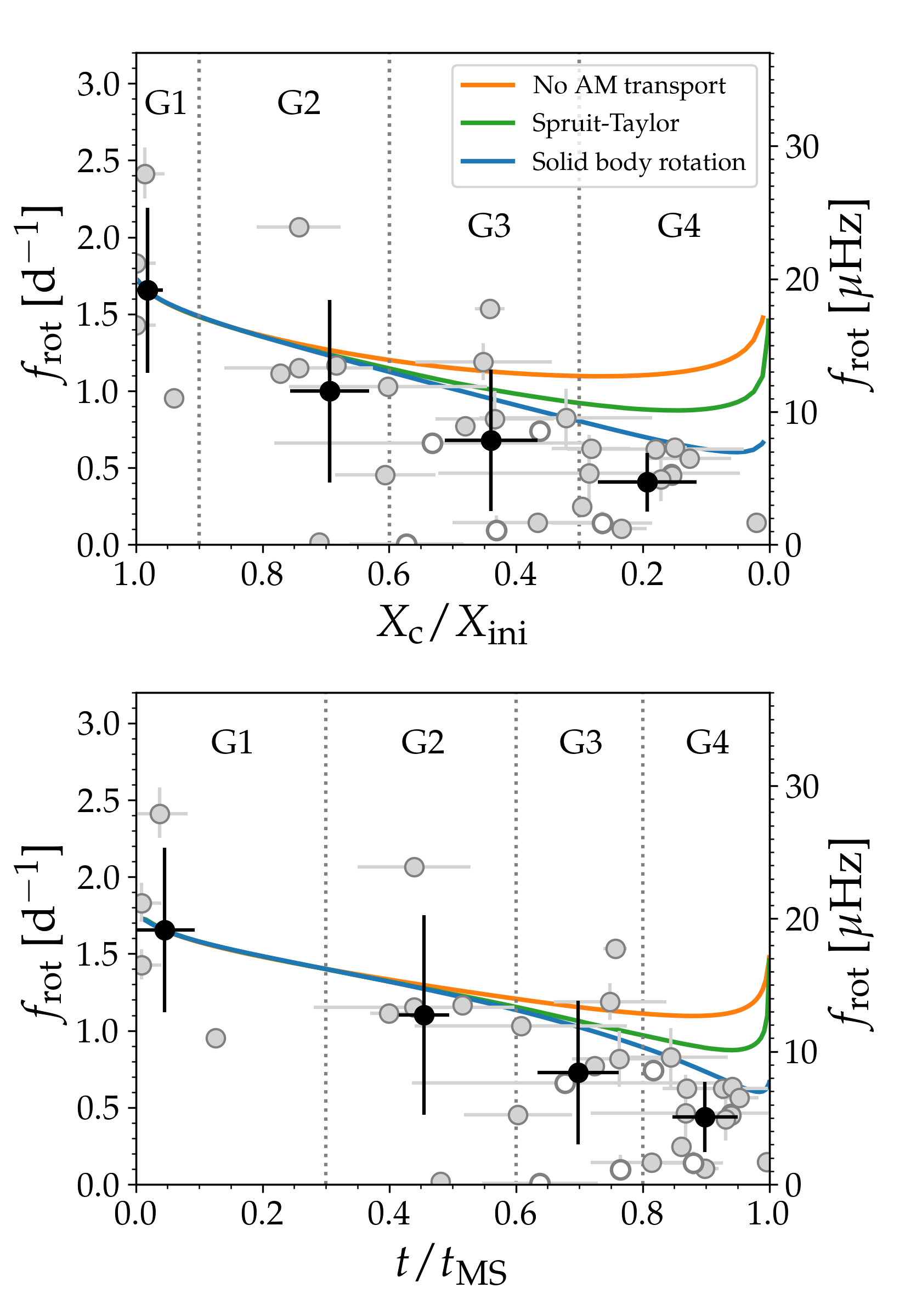}
\caption{Distribution of near-core rotation frequencies with age for the 33 SPB stars with space photometry (grey). Binary stars are indicated by open symbols. The vertical dotted lines show the dividing ages for the four different age groups G1-G4, and the black data points give the position of the average rotation and ages of each age group. The evolution of the near-core rotation for the case of rigid rotation (blue) and no angular momentum transport (orange) has been scaled to the average rotation and age in group G1. The same results assuming models with angular momentum transport from the Spruit-Taylor dynamo are shown in green for comparison.}
	\label{fig:AM_space}
\end{figure}

We split the 33 SPB stars with space photometry into four age groups and calculate the average near-core rotation frequency and age for the sample in each group. For this analysis we focus on the two age indicators $X_{\rm c}/X_{\rm ini}$ and $t/t_{\rm MS}$. The four different age groups are:

\begin{itemize}
	\item[G1:] $X_{\rm c}/X_{\rm ini} > 0.9$ and $t/t_{\rm MS} < 0.3$,
	\item[G2:] $0.6 < X_{\rm c}/X_{\rm ini} < 0.9$ and $ 0.3 < t/t_{\rm MS} < 0.6$,
	\item[G3:] $0.3 < X_{\rm c}/X_{\rm ini} < 0.6$ and $ 0.6 < t/t_{\rm MS} < 0.8$,
	\item[G4:] $X_{\rm c}/X_{\rm ini} < 0.3$ and $t/t_{\rm MS} > 0.8$.
\end{itemize}

We show again the distribution of rotational frequencies as a function of age in Fig.~\ref{fig:AM_space}, where the vertical dotted lines separate the four age groups. The averages of the age groups and their $1\sigma_{\rm std}$ are shown in black.

Using the curves for the evolution of the near-core rotation frequency shown for the 5\,M$_\odot$ star in the left-hand side of Fig.~\ref{fig:frot_cor_surf}, we scale the curves to the average rotation frequency of the four youngest SPB stars and compare the expected evolution of the near-core rotation with the averages in each of the three older age groups. For both G2 and G3 we find that the observed averages are consistent with the predicted evolution of the near-core rotation frequency for both the case of no and highly efficient angular momentum transport. However, for the oldest age group G4, the predicted rotation rate for the sample is 1.72 and 1.66 times higher than the observed average for the case of highly efficient angular momentum transport for $X_{\rm c}/X_{\rm ini}$ and $t/t_{\rm MS}$, respectively, but within $1.54\,\sigma_{\rm std}$ and $1.28\,\sigma_{\rm std}$ of the average. The discrepancy is even higher for the case of no angular momentum transport. Here the predicted rotation frequency is 2.7 and 2.5 times higher for $X_{\rm c}/X_{\rm ini}$ and $t/t_{\rm MS}$, respectively, corresponding to a $3.65\,\sigma_{\rm std}$ and $2.89\,\sigma_{\rm std}$ deviation from the observed average. Furthermore, we note that none of the SPB stars are found within the gap between the two curves in the age range of G4. This is seen most clearly for the $X_{\rm c}/X_{\rm ini}$ age indicator.

\begin{figure*}
\includegraphics[width=0.5\linewidth]{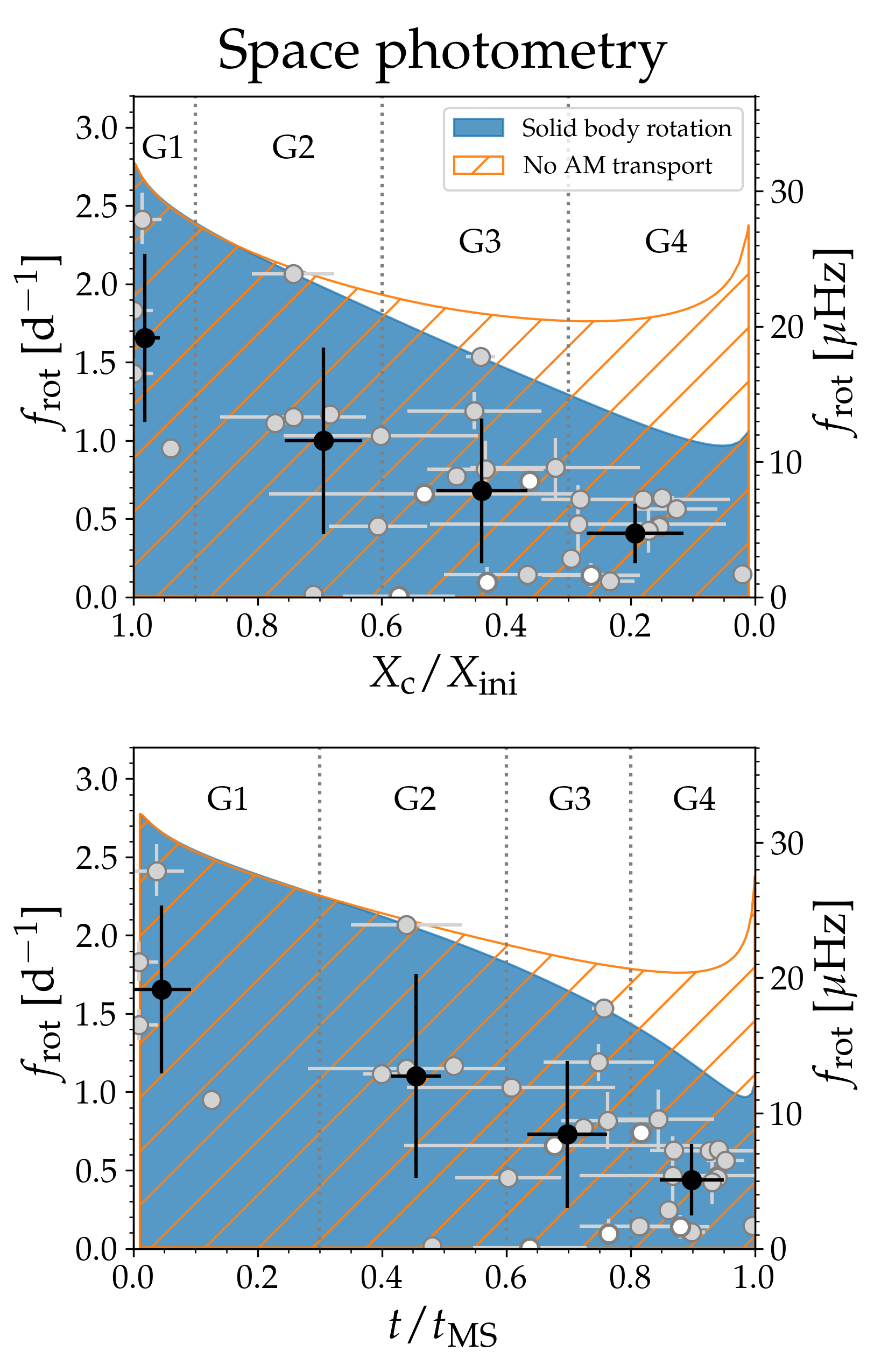}
\includegraphics[width=0.5\linewidth]{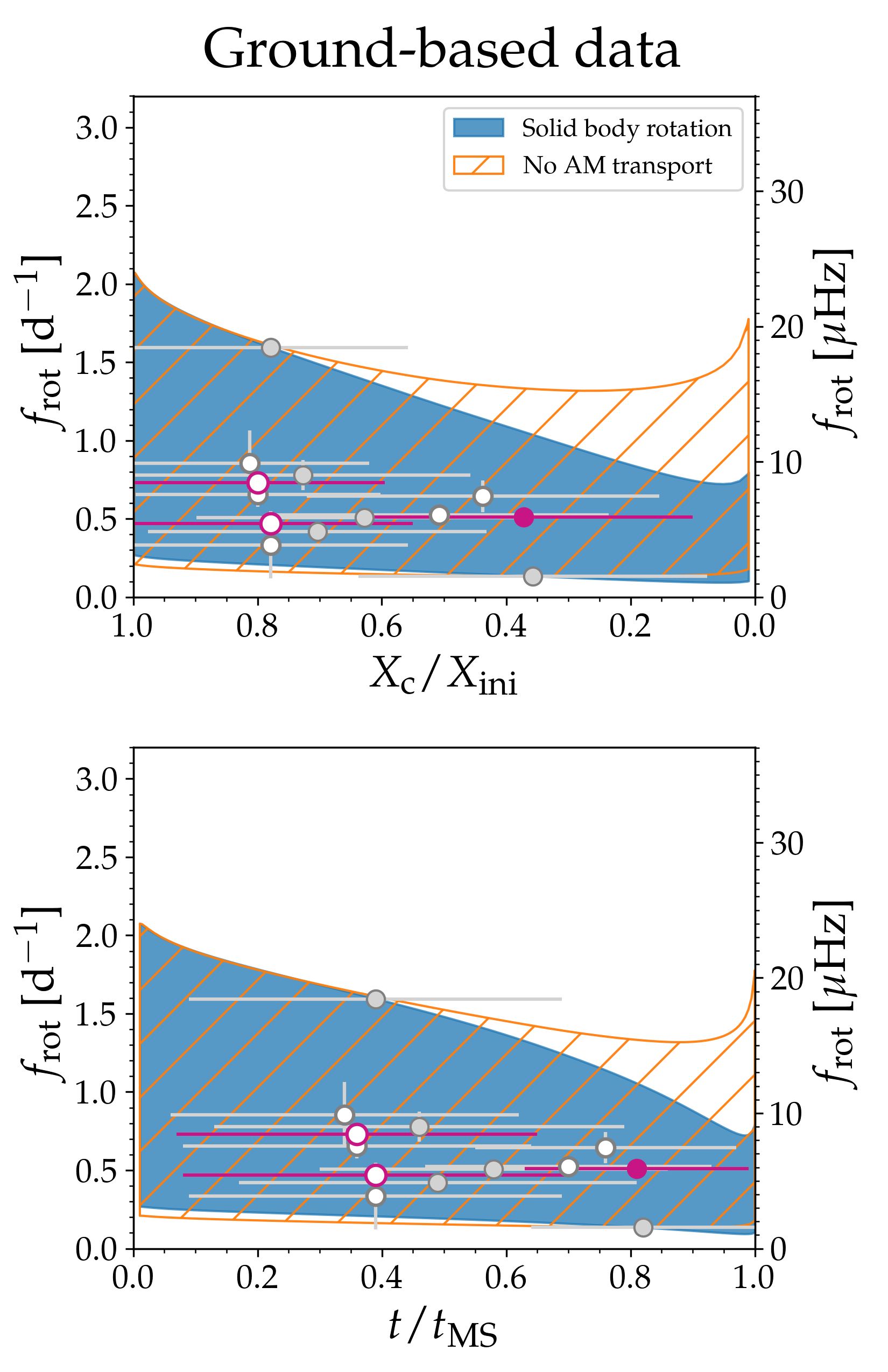}
\caption{Expected ranges of observed near-core rotation frequencies as predicted for the sample of SPB stars assuming rigid rotation (blue) and no angular momentum transport (hatched orange). The predicted regions are shown separately for the 33 stars with space photometry (left) and 13 stars with ground based data (right) which have measured $X_{\rm c}/X_{\rm ini}$ and $t/t_{\rm MS}$ values. SPB stars which show and inconsistency between the derived inclination angles and mode identification are shown in purple. See Fig.~\ref{fig:AM_space} for further explanations of the figure details.}
	\label{fig:AM_regions}
\end{figure*}

Next we take a closer look at this gap and derive the regions for the distribution of the near-core rotation with age spanned by the sample of SPB stars assuming no and highly efficient angular momentum transport. We do these calculations separately for the sample with space photometry and ground-based data. The regions are derived by scaling the same curves for the 5\,M$_\odot$ star in Fig.~\ref{fig:frot_cor_surf} to each of the 46 SPB stars with a known value for $X_{\rm c}/X_{\rm ini}$ and $t/t_{\rm MS}$, and then taking the minimum and maximum predicted rotation at each age step. The resulting regions are shown in Fig.~\ref{fig:AM_regions}. As seen in the figure, we end up with an area spanned by the calculations without angular momentum transport (hatched orange) where old SPB stars are expected to be found, but are absent in the case of solid body rotation (full blue). None of the 46 SPB stars are found in this area of the diagrams, implying that the sample shows evidence of efficient angular momentum transport on the main-sequence.

\section{Inclinations}\label{Sec:Inclination}

\begin{figure}
\includegraphics[width=\linewidth]{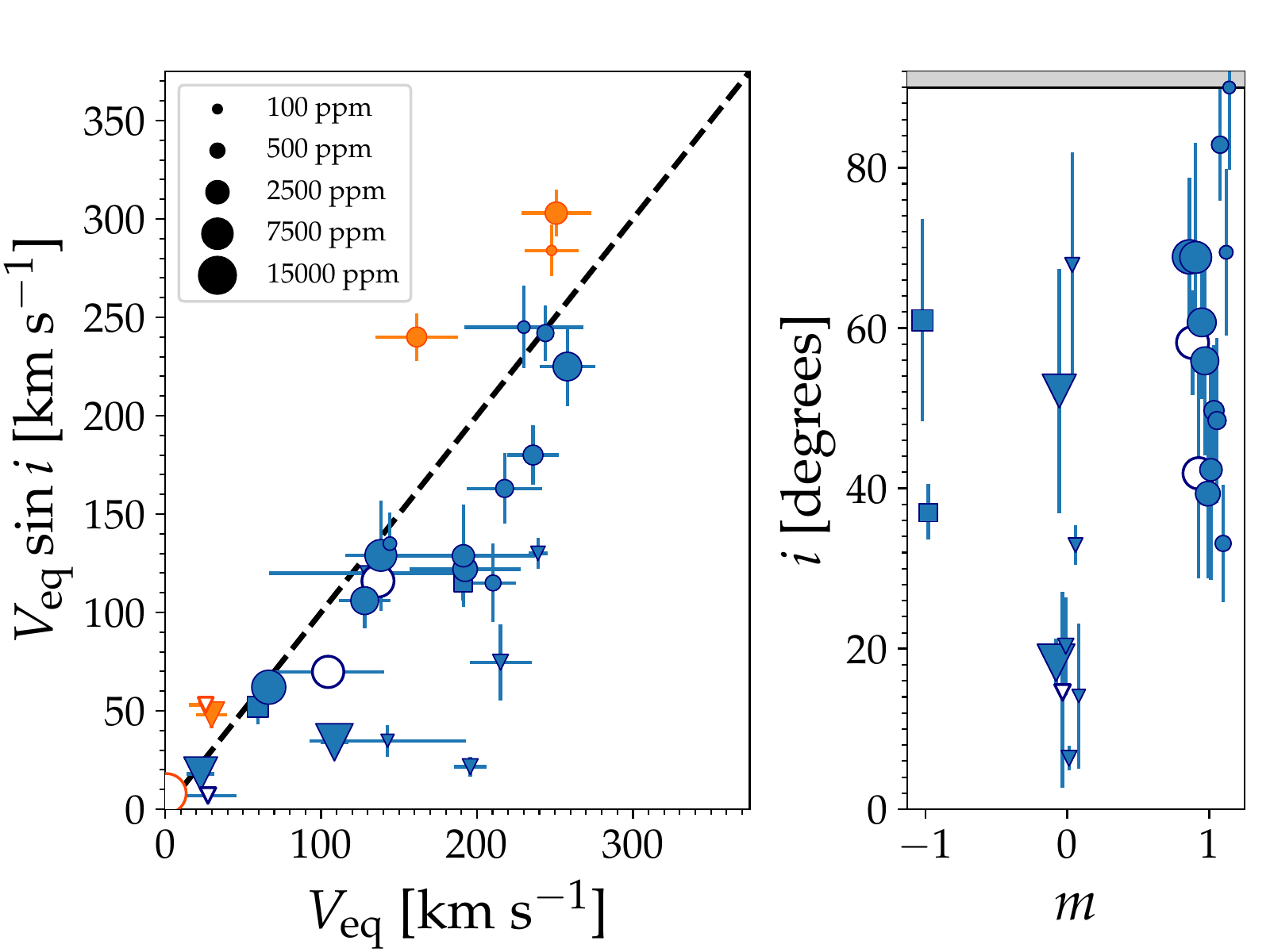}
\caption{\emph{Left:} Spectroscopic $V_{\rm eq} \sin i$ as a function of the asteroseismic $V_{\rm eq}$ derived under the assumption of rigid rotation for the sample of SPB stars with space photometry. The size of the data points scale with the amplitude of the dominant dipole mode, while the symbols indicate whether the oscillations are retrograde (square), zonal (triangle), or prograde (circles) modes. Orange data points correspond to stars for which the $V_{\rm eq} \sin i$ cannot be reconciled with the $V_{\rm eq}$ values within the estimated errors. \emph{Right:} Derived inclination angles as a function of the azimuthal order. The data points have been slightly offset from the measured $m$ value for the sake of visibility. The black horizontal line marks where $i=90^{\rm o}$, i.e. the maximum inclination angle. Binary stars are indicated with open symbols.}
	\label{fig:inclination_space}
\end{figure}

\begin{figure}
\includegraphics[width=\linewidth]{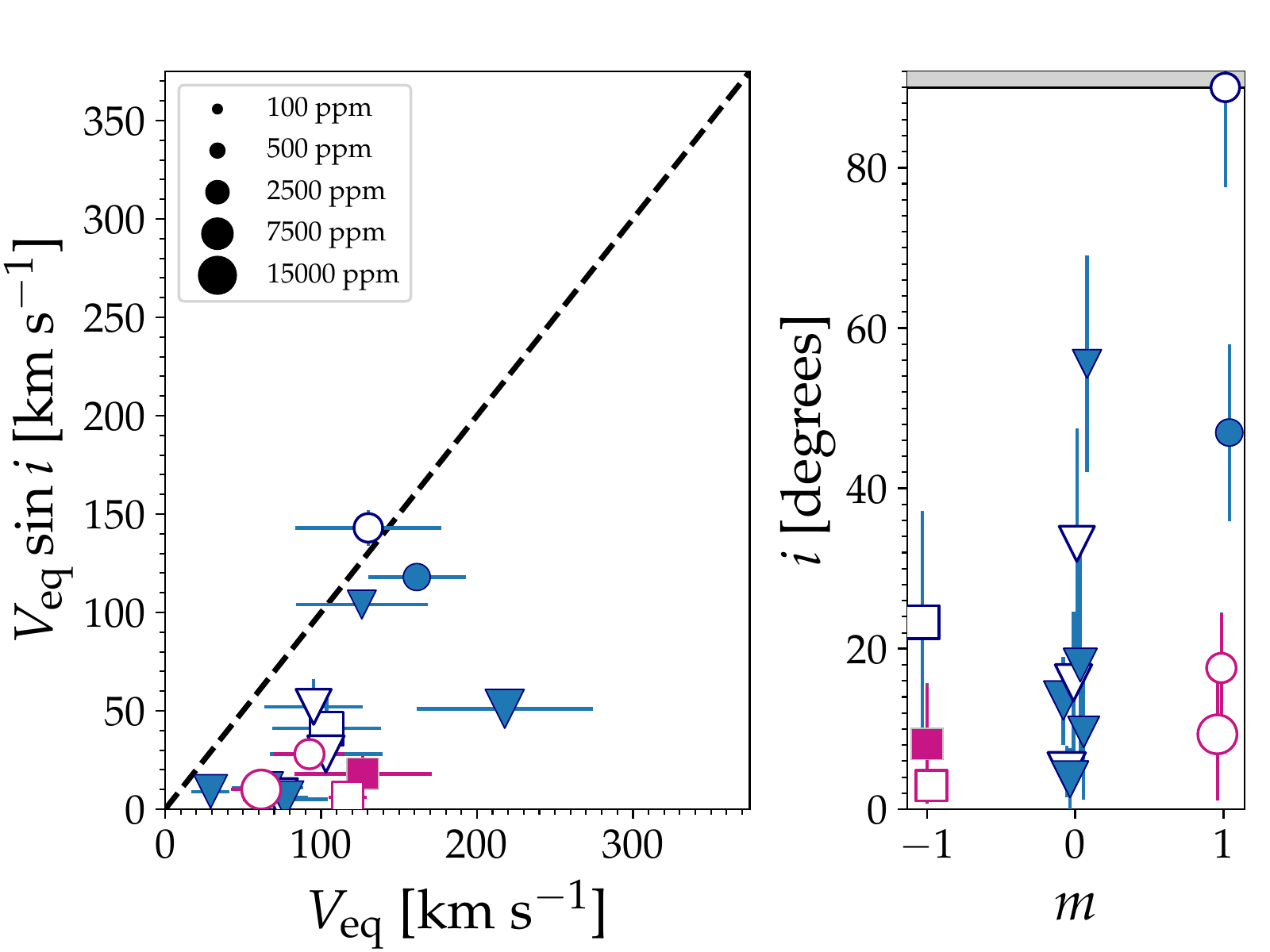}
\caption{Same as Fig. \ref{fig:inclination_space} but for SPB stars observed from the ground. Stars with $i < 20^{\rm o}$ are indicated in purple.}
	\label{fig:inclination_ground}
\end{figure}

\noindent Under the assumption of uniform rotation, we can estimate the inclination angles $i$ of the stars if their $V_{\rm eq} \sin i$ values have been measured spectroscopically. This is the case for 46 out of the 52 SPB stars. As shown in Eq.~(\ref{Eq:Veq}) the measured rotation frequencies and stellar radii directly give us the equatorial velocity of a spherical symmetric star. The adopted $V_{\rm eq} \sin i$ values and stellar radii $R$ are listed in Table~\ref{Tab:rot_parameters} along with the derived inclination angles. We show the results in Figs.~\ref{fig:inclination_space} and \ref{fig:inclination_ground} for the dipole modes of the SPB stars with space photometry and ground-based data, respectively, and note that the assumption of uniform rotation matches well with what has previously been observed for single main-sequence stars with convective cores for which both asteroseismic core and envelope rotation frequencies have been measured, as illustrated in Fig.~4 of \cite{Aerts2019b}.

The left hand sides of Figs.~\ref{fig:inclination_space} and \ref{fig:inclination_ground} compare the observed spectroscopic $V_{\rm eq} \sin i$ to the asteroseismically derived $V_{\rm eq}$. As the $V_{\rm eq} \sin i$ value is only a lower estimate of the true $V_{\rm eq}$ value, we would expect to always find $V_{\rm eq} \sin i \leq V_{\rm eq}$. However, as highlighted by the orange symbols there are six stars (KIC~4936089, KIC~8264293, KIC~8459899, KIC~8766405, KIC~11360704, HD 201433) in Fig.~\ref{fig:inclination_space} for which the asteroseismic equatorial velocity cannot be reconciled with the measured $V_{\rm eq} \sin i$ values within the $1\sigma$ errors. There are a few reasons why this would be the case. For one, we have assumed rigid rotation for the derivation of $V_{\rm eq}$, which might not be valid for these stars. In this case, the envelopes of the stars would have to spin faster than the cores in order to be able to reconcile the spectroscopic $V_{\rm eq} \sin i$ with the asteroseismic $V_{\rm eq}$. 
This is the case for the SPB star HD~201433, which is a known single-lined spectroscopic triple system. Using nine observed rotationally split dipole g-modes, \cite{Kallinger2017} derived the internal rotation profile of the star showing that it is rigidly rotating up to $\sim 90\%$ of the stellar radius, but the surface (i.e. outer $\sim 4\%$) has spun up likely due to tidal interactions with its inner companion. Based on visibility arguments of the components of the rotationally split g-modes, \cite{Kallinger2017} arrived at an inclination of $68\pm 5^{\rm o}$ for HD~201433. For this to be consistent with the observed $V_{\rm eq} \sin i$ would require $V_{\rm eq} \approx 8.6$\,km\,s$^{-1}$ corresponding to $f_{\rm rot} \approx 0.0615$\,d$^{-1}$ at the surface. 

On the other hand, an envelope rotating faster than the core has previously been measured for a main-sequence $\gamma$~Dor star showing both g- and p-mode oscillations \citep{Kurtz2014}. Similarly, the internal rotation profile of KIC\,10526294 has previously been inferred by \cite{Triana2015}, who showed that not only is the envelope rotating faster than the core of the star but it is also rotating in the opposite direction. Internal gravity waves (IGWs) generated by the convective core have been proposed as an efficient mechanism to transport angular momentum \citep{Rogers2013}, possibly explaining both of these observed rotation profiles \citep{Rogers2015}.

Another possible explanation is related to the fact that the total broadening of the spectral lines used to measure the $V_{\rm eq} \sin i$ value is not only affected by the rotation of the stars, but also by stellar oscillations which give rise to a time dependent pulsational broadening of the line-profiles \citep{Aerts2014}. As a result, the $V_{\rm eq} \sin i$ will vary as a function of time, and its measured value is a sum of the contribution from the rotational and pulsational broadening. In other words, the measured spectroscopic $V_{\rm eq} \sin i$ for an oscillating star is an upper limit of the true $V_{\rm eq} \sin i$ value. The effects of pulsational broadening on the estimated $V_{\rm eq} \sin i$ is expected to be larger for slowly rotating stars \citep{Aerts2014}.

To investigate any possible correlations with the pulsations on the measured $V_{\rm eq} \sin i$, the symbol sizes in Figs.~\ref{fig:inclination_space} and \ref{fig:inclination_ground} have been scaled according to the amplitude of the dominant g-mode with a known mode identification for each star. For stars with period spacing patterns, the dominant mode of the pattern used to derive the near-core rotation frequency was used. As the photometric data come from a variety of different telescopes, we converted all amplitudes to the Johnson V band and units of parts-per-million as described in detail in Appendix~\ref{App:Amp_corr}.
 As seen in Figs.~\ref{fig:inclination_space} and \ref{fig:inclination_ground}, the pulsations of the slowest rotating stars tend to have higher amplitudes. This is also seen in Fig.~\ref{fig:inclination_ground}, where the considered SPB stars with ground-based data cover a smaller range in surface velocities and have generally higher amplitudes than the sample of SPB stars with space photometry. This could explain the lack of observed fast rotating SPB stars from ground-based observations \citep[e.g.][]{DeCat2002}. 

The right hand panels in Figs.~\ref{fig:inclination_space} and \ref{fig:inclination_ground} show the derived inclination angles $i$ for the 40 SPB stars with measured $V_{\rm eq} \sin i < V_{\rm eq}$, as a function of the azimuthal order including a small offset for the sake of clarity. Focusing first on Fig.~\ref{fig:inclination_space}, we see that the stars with zonal modes tend to have smaller inclination angles than the stars with either retrograde or prograde modes. This is entirely in line with the theoretical expectations from surface cancellation effects should there be equipartition among the components of multiplets, as for solar-like oscillators \citep[][see Fig.~2]{Gizon2003}. Although there is no equipartition of mode energy among multiplet components of heat-driven modes, this is exactly what is seen in Fig.~\ref{fig:inclination_space}.

The results are slightly different for the sample of SPB stars with ground-based data as shown in Fig.~\ref{fig:inclination_ground}. Here four stars (HD~24587, HD~85953, HD~160762, HD~182255) with either prograde or retrograde modes are found to have $i < 20^{\rm o}$. There are a few reasons why such low inclination angles might be obtained. The first is that the estimated stellar parameters might be wrong. As an example, changing the stellar radius of HD~85953 from $R=4.9\pm1.7$\,R$_\odot$ \citep[][adapted in this work]{Hubrig2006} to $R=4.7\pm1.0$\,R$_\odot$ \citep{DeCat2002} increases the inclination of the star to $44\pm 17^{\rm o}$ and in line with the trend that is seen for the SPB stars with space photometry. Given that the stellar parameters of the sample of SPB stars with ground based data included in Fig.~\ref{fig:inclination_ground} have been derived by comparing the position of the stars in the HR diagram with evolutionary tracks, much better parameter estimates could be obtained in the future if asteroseismic modeling is accomplished for the stars.
Another possibility is that mode identification is wrong, and that the dipole prograde/retrograde oscillations with $i < 20^{\rm o}$ are actually zonal modes.

Finally, a third possibility is that the assumption of rigid rotation is inaccurate and the stars are differentially rotating in their radiative envelopes. In this case, the surface of the stars would have to be rotating slower than their near-core regions to explain the inclination angles. Increasing the inclination above $20^{\rm o}$ would require the surface to be rotating at least 1.1, 2.4, 6.6, and 2.1 times slower than the near-core regions for HD~24587, HD~85953, HD~160762 and HD~182255, respectively, assuming all other relevant parameters stay the same. Future asteroseismic modeling of these stars combining ground-based data with data from the TESS space telescope hold the promise of bringing answers to these questions as well as significantly improving the age estimates for the sample of SPB stars with ground-based data. 

\section{Conclusions}\label{sec:Conclusions}

\noindent In this work, we took a closer look at the internal rotation properties of 52 SPB stars which have an asteroseismic measurement of the internal rotation based on their g-mode oscillations. The spin parameters of the stars reveal that most of the g-modes are located in the sub-inertial regime, where the effects of the Coriolis force cannot be treated as a simple perturbation to the oscillation equations \citep{Ballot2010}. This was taken into account in the initial calculation of the rotation frequency for the majority of the stars in the sample through the use of the TAR, in which the centrifugal force is ignored. \cite{Henneco2021} demonstrated that the detectability of the effect of the centrifugal force, following the extension of the TAR by \cite{Mathis2019}, increases with increasing rotation rate but decreases as the stars become older. As the frequency resolution is equal to the inverse of the length of the light curve, they also showed that the effects are easier to detect for stars observed by \emph{Kepler} with its four years of continuous observations, compared to stars observed in the continuous viewing zone (351\,d) of the \emph{TESS} space telescope \citep{Ricker2015}.

Independent of the considered age indicator, the near-core rotation frequencies  are found to decrease as a function of the main-sequence evolution for the 52 SPB stars. The only exception is seen for the sample of stars whose asteroseismic analysis was based on data from ground-based telescopes, where no correlation is found between $f_{\rm rot}$ and $\log g$. We speculate that this might be caused by differences in the adopted method for deriving the internal rotation for three of the slowest rotating stars in the sample. 

Next, we tested if the decrease in near-core rotation rate is consistent with the expectations for the evolution of the internal rotation profile in the two limiting cases of no and highly efficient angular momentum transport. We find that both descriptions predict that the near-core rotation frequency decreases during the main-sequence evolution but at different rates. Calculations without angular momentum transport predict higher rotation frequencies towards the end of the main-sequence evolution than the case of highly efficient angular momentum transport corresponding to rigid rotation, which is inconsistent with the observations for the 52 SPB star sample. Therefore, the lack of old SPB stars with rapid near-core rotation rates is evidence of angular momentum transport occurring throughout the main-sequence evolution of the stars.

Finally, we use the assumption of rigid rotation to derive the inclination angles of the stars for which spectroscopic $V_{\rm eq} \sin i$ values are available. For the sample of SPB stars with space photometry, we find that the stars with zonal modes tend to have smaller inclination angles ($i \leq 70^{\rm o}$), while stars with prograde or retrograde modes have higher inclination angles ($i \geq 30^{\rm o}$). This is consistent with the expectations from surface cancellation effects. Four stars with ground-based data, however, show $m=-1$ and $m=1$ modes at low inclination angles ($i \leq 20^{\rm o}$). Future asteroseismic modeling of the stars will be crucial for revealing the reason for these low inclination angles.

We note that the results presented here are currently limited by the sample size. Most importantly, the sample has a low fraction of young SPB stars with only two out of the 52 SPB stars being known to be zero-age main-sequence stars \citep{Zwintz2017}. Increasing the number of young SPB stars with near-core rotation measured from asteroseismology will be crucial for anchoring and testing models with different levels of angular momentum transport such as previously done for $\gamma$~Dor stars by \cite{Ouazzani2019}.

Furthermore, while the current sample of SPB stars show evidence of efficient angular momentum transport occurring throughout the main-sequence evolution, it does not currently provide information on the level of internal differential rotation in the stars. The only exceptions are two of slowly rotating SPB stars in the sample, where inversion techniques were previously applied to the g-mode oscillations to derive the internal rotation profile \citep{Triana2015,Kallinger2017}. Improvements can be made by combining near-core rotation measurements with surface rotation measurements from either rotationally split p-modes or rotational spot modulation. However, period spacing patterns on their own can provide important information about the internal rotation profile of the star. \cite{VanReeth2018} showed that the level of internal differential rotation can be measured if more than one period spacing pattern is detected in the stars, which is the case for several of the SPB stars in the current sample \citep[see e.g.][]{Pedersen2020PhD,Szewczuk2021}. Finally, \cite{Ouazzani2020} demonstrated that the coupling of pure inertial modes in the convective core with gravito-inertial modes in the radiative envelope give rise to unique features in the period spacing patterns, which can be used to derive the rotation frequency of the convective core. This was recently accomplished for 16 $\gamma$~Dor stars by \cite{Saio2021}, who found the majority of the stars to be rigidly rotating while a few of the less evolved $\gamma$~Dor stars have convective cores rotating slightly faster than the envelope in the g-mode propagation cavity. Such measurements have yet to be accomplished for SPB stars.\\


{\small
\noindent The author is thankful to Conny Aerts, Lars Bildsten, Jim Fuller, and Jamie Tayar for providing interesting discussions and useful comments at different stages of this work. The author is also thankful to the anonymous referee whose comments improved the manuscript. This research was supported in part by the National Science Foundation under Grant No. NSF PHY-1748958. It was performed in part at Aspen Center for Physics, which is supported by National Science Foundation grant PHY-1607611, and was partially supported by a grant from the Simons Foundation. This research has made use of the SIMBAD database,
operated at CDS, Strasbourg, France, and of the VizieR catalogue access tool, CDS, Strasbourg, France (DOI: 10.26093/cds/vizier). The original description of the VizieR service was published in A\&AS 143, 23. This research has made use of the Spanish Virtual Observatory (https://svo.cab.inta-csic.es) project funded by MCIN/AEI/10.13039/501100011033/ through grant PID2020-112949GB-I00
}

%

\vspace{15mm}




\appendix


\section{Age indicators}\label{App:Age_indicators}

This work considers three different age indicators to study the evolution of the near-core rotation as a function of age on the main-sequence. As a minimum we require that the surface gravity $\log g$ has to be known, which is the age parameter previously used by \cite{Aerts2019b,Aerts2021} to study the evolution of the near-core rotation frequency from the main-sequence to the white dwarf evolutionary stage. We argue, however, that this is the least accurate age parameter out of the three that are considered in this work as the surface gravity not only depends on age, but also the mass, internal mixing, and metallicity of the stars. The two remaining age indicators are the ratio of the current to initial core hydrogen mass fraction $X_{\rm c}/ X_{\rm ini}$, and the ratio of the current age to the main-sequence lifetime $t/t_{\rm MS}$. By considering the ratios instead of $X_{\rm c}$ and $t$, we minimize the dependence of mass, metallicity, and internal mixing on the age indicators as we will see below.

To make a comparison between the three age indicators at different main-sequence ages and stellar masses, we adopt the same \texttt{MESA} setup as in \cite{Pedersen2021} and compute evolutionary tracks for four different stellar masses: $2.5$\,M$_\odot$, $5$\,M$_\odot$, $7.5$\,M$_\odot$, and $10$\,M$_\odot$. The initial metallicity is set to $Z = 0.014$ and we adopt exponential diffusive overshoot and constant envelope mixing for the internal mixing profile, corresponding to mixing profile labeled $\boldsymbol{\psi_1}$ by \cite{Pedersen2021}. For the mixing parameters we choose $f_{\rm ov} = 0.01$ and $D_{\rm env, 0} = 10$\,cm$^2$\,s$^{-1}$, where the former sets the extent of the convective core boundary mixing region and the second the level of mixing in the radiative envelope \citep[see][for further details]{Pedersen2021}. We use the same set of computed stellar models for the calculations of the evolution of the near-core rotation with age in Sect.~\ref{Sec:AM}. To make sure that the total angular momentum is conserved for these calculations, we excluded stellar winds from the \texttt{MESA} calculations. We find that doing so does not significantly impact the results presented in this work.

\begin{figure}
\center
\includegraphics[width=\linewidth]{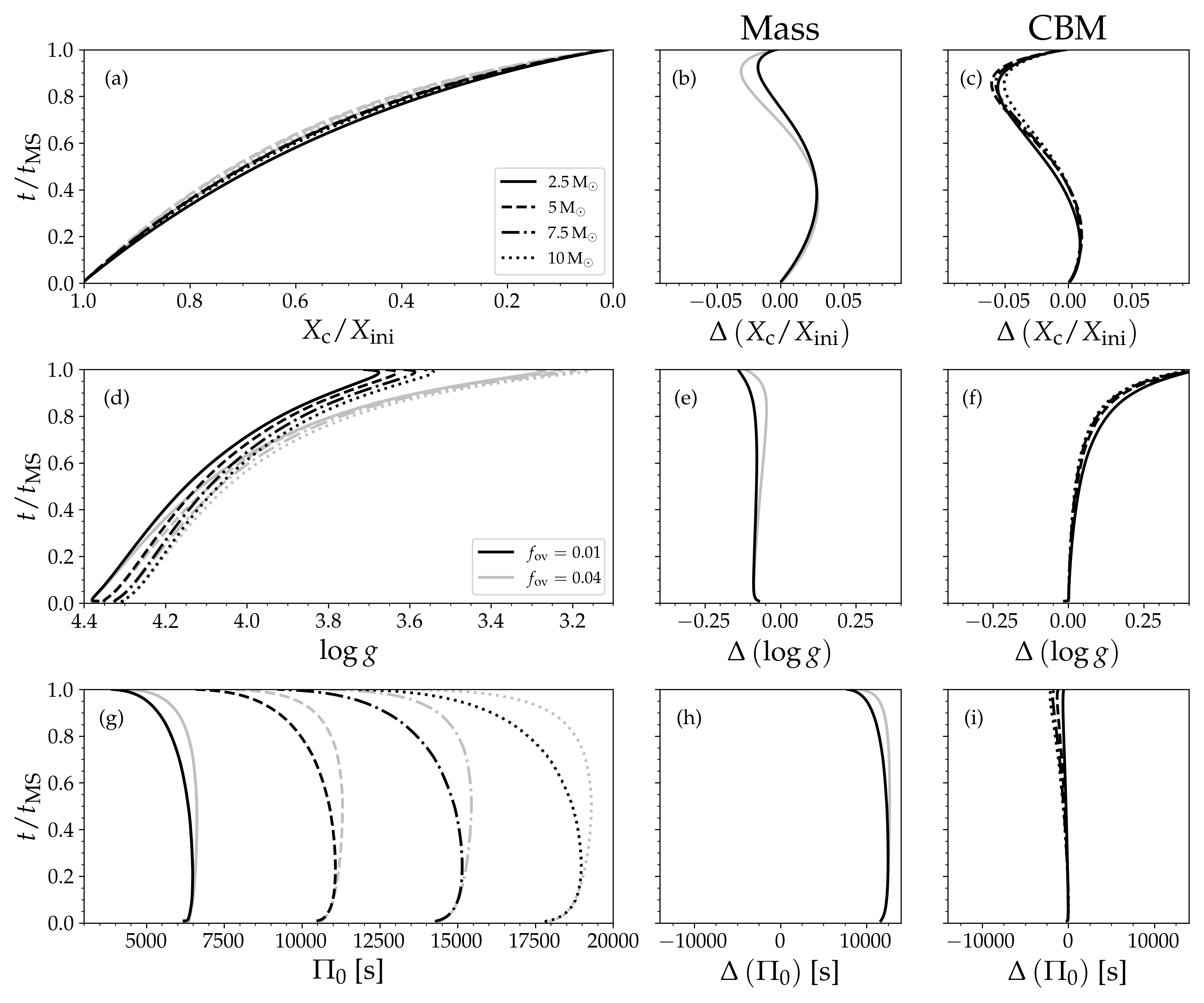}
\caption{Comparison between four different age indicators $t/t_{\rm MS}$, $X_{\rm c}/ X_{\rm ini}$, $\log g$, and $\Pi_0$ for different masses and exponential diffusive overshooting. Panels (a), (c), and (g) show $t/t_{\rm MS}$ as a function of $X_{\rm c}/ X_{\rm ini}$, $\log g$, and $\Pi_0$, respectively. Different masses correspond to different line styles: 2.5\,M$_\odot$ (full), 5\,M$_\odot$ (dashed), 7.5\,M$_\odot$ (dot-dashed), and 10\,M$_\odot$ (dotted). Black lines correspond to models with $f_{\rm ov} = 0.01$, whereas models with $f_{\rm ov} = 0.04$ are shown in grey. Panels (b), (e), and (h) show the differences in $X_{\rm c}/ X_{\rm ini}$, $\log g$, and $\Pi_0$ obtained for a given $t/t_{\rm MS}$ between the highest and lowest mass model, e.g. $\Delta \left( X_{\rm c} / X_{\rm ini} \right) = \left.X_{\rm c} / X_{\rm ini}\right\vert_{10\,{\rm M}_\odot} - \left.X_{\rm c} / X_{\rm ini}\right\vert_{2.5\,{\rm M}_\odot}$. Panels (c), (f), and (i) gives the same differences but between same mass models with different overshooting parameters, e.g. $\Delta \left( X_{\rm c} / X_{\rm ini} \right) = \left.X_{\rm c} / X_{\rm ini}\right\vert_{f_{\rm ov}=0.01} - \left.X_{\rm c} / X_{\rm ini}\right\vert_{f_{\rm ov}=0.04}$.}
	\label{fig:age_ind}
\end{figure}

The top two rows of Fig.~\ref{fig:age_ind} show $t/t_{\rm MS}$ as a function of $X_{\rm c}/ X_{\rm ini}$ (panel a) and $\log g$ (panel d), respectively, for the main-sequence evolution of the stars. The differences in line styles corresponds to different stellar masses, whereas the grey curves show the same calculations for a larger convective core boundary mixing region ($f_{\rm ov} = 0.04$). Focusing first on panel (a), the relation between $t/t_{\rm MS}$ and $X_{\rm c}/ X_{\rm ini}$ is not purely linear, as the stellar evolution speeds up towards hydrogen depletion. The relations are nearly the same independently of the mass of the star and the extent of the convective core boundary mixing region, but models with higher overshooting tend to have higher $t/t_{\rm MS}$ values for the same $X_{\rm c}/ X_{\rm ini}$. This is seen more clearly in panels (b) and (c), which show the difference in $X_{\rm c}/ X_{\rm ini}$ as a function of $t/t_{\rm MS}$ obtained between the highest and lowest mass models (panel b) and the models with the largest and smallest extent of the convective core boundary mixing region (panel c). Increasing the overshooting parameter leads to larger differences in $X_{\rm c}/ X_{\rm ini}$ compared to those obtained by increasing the stellar mass, but in both cases the differences remain smaller than $\approx 0.05$.

In comparison, increasing the stellar mass from 2.5\,M$_\odot$ to 10\,M$_\odot$ gives rise to larger differences in $t/t_{\rm MS}$ for the same value of $\log g$. Panel (d) shows that increasing the stellar mass causes the $t/t_{\rm MS}$ versus $\log g$ curves to shift towards higher $\log g$ values, as also implied in panel (e). Furthermore, models with higher overshooting starts to deviate from those with small convective core boundary mixing regions for $t/t_{\rm MS} \gtrsim 0.2$ for the same stellar mass. The differences increases for increasing $t/t_{\rm MS}$ and become larger than $\approx$0.4\,dex at the end of the main-sequence evolution (see panel (f)). This is why we consider $\log g$ to be the least accurate age indicator out the three considered in this work.

In their study of internal angular momentum transport using a sample of 37 $\gamma$~Dor stars, \cite{Ouazzani2019} used the bouyancy radius, also known as the asymptotic period spacing

\begin{equation}
\Pi_0 = 2 \pi^2 \left( \int_{\rm gc} \frac{N}{r} {\rm d}r \right)^{-1},
\end{equation}

as the age indicator. They demonstrated that $\Pi_0$ is monotonously decreasing as a function of $X_{\rm c}/X_{\rm ini}$ for their considered mass range of $1.4-1.8$\,M$_\odot$ and differs by up to $\approx 800\,{\rm s}$ for a given $X_{\rm c}/X_{\rm ini}$ \citep[][see their Fig.~2]{Ouazzani2019} . In comparison, $\Pi_0$ on its own is an unreliable age indicator for SPB stars as demonstrated in panels (g)-(i) in Fig.~\ref{fig:age_ind}. There are multiple reasons for this. First of all, $\Pi_0$ is largely dependent of the stellar mass. As the SPB stars span a much larger range in stellar masses than the $\gamma$~Dor stars, this correspondingly leads to a much larger difference in $\Pi_0$ for the same value of $t/t_{\rm MS}$ as seen in panel (h) when comparing the 2.5\,M$_\odot$ to the 10\,M$_\odot$ models which results in $\Delta (\Pi_0) > 10000$\,s. This difference is generally much larger than the intrinsic variation in $\Pi_0$ for a given mass throughout the main-sequence evolution. Secondly, for SPB stars the $\Pi_0$ values are no longer monotonously decreasing with age. Instead, $\Pi_0$ is increasing in the beginning of the main-sequence evolution and later starts to decrease for the rest of the evolution. Increasing the extent of the convective core boundary mixing region causes the $\Pi_0$ values to stay nearly the same or increasing for a larger part of the main-sequence evolution before rapidly decreasing at $t/t_{\rm MS} \gtrsim 0.8$, see also panel (i). For these reasons, we chose not to consider the evolution of $f_{\rm rot}$ as a function of $\Pi_0$ in this work.

While all of the considered 52 SPB stars have a known $\log g$ value either from spectroscopy, multi-color Str{\"o}mgren photometry, or asteroseismic modeling, the $X_{\rm c}/ X_{\rm ini}$ and $t/t_{\rm MS}$ values tend not to be given simultaneously for the same star. Therefore, we choose to do a conversion between $X_{\rm c}/ X_{\rm ini}$ and $t/t_{\rm MS}$ and vise versa by interpolating onto the curve of the 5\,M$_\odot$ star with $f_{\rm ov} = 0.01$ in Fig.~\ref{fig:age_ind} (a).


\section{MESA models including the Spruit-Taylor dynamo}\label{App:MESA_ST}

In oder to compare the expected evolution of the near-core and surface rotation assuming rigid rotation throughout the main-sequence evolution and no angular momentum transport to the predictions when angular momentum transport occurs through the Spruit-Taylor dynamo (shown in green in Figs.~\ref{fig:frot_cor_surf} and \ref{fig:AM_space}), we use the same \texttt{MESA} setup discussed in Appendix~\ref{App:Age_indicators} but with a few modifications.

The evolution of the stellar models is split into two, with the first being the pre-main-sequence evolution where no angular momentum transport is enforced. The evolution is stopped close to the zero-age main-sequence once $L_{\rm nuc}/L \geq 0.95$ is fulfilled, where $L_{\rm nuc}$ is the total luminosity from nuclear reactions, and a starting model for the main-sequence evolution is saved. In the second step, the saved model is loaded into \texttt{MESA} and evolved until the core hydrogen mass fraction drops below 0.001. For the main-sequence evolution, the starting model is assumed to be rigidly rotating at $10\,\%$ of the critical rotation rate for the green curves shown in Figs.~\ref{fig:frot_cor_surf} and \ref{fig:AM_space}. The Spruit-Taylor dynamo is assumed to be the only contributor to the angular momentum transport and internal element mixing from rotation is turned off. The latter is to make sure that the adopted angular momentum transport mechanism is the only difference between these models and the computed models described in Appendix~\ref{App:Age_indicators}. The corresponding additional \texttt{MESA} \texttt{inlist} parameters adopted for the main-sequence evolution are\\

\noindent\texttt{change\_v\_flag = .true.}\\
\texttt{new\_v\_flag = .true.}\\

\noindent\texttt{change\_rotation\_flag= .true.}\\
\texttt{new\_rotation\_flag = .true.}\\

\noindent\texttt{new\_omega\_div\_omega\_crit = 0.1}\\
\texttt{relax\_omega\_div\_omega\_crit = .true.}\\
\texttt{relax\_initial\_omega\_div\_omega\_crit = .true.}\\

\noindent\texttt{change\_D\_omega\_flag = .true.}\\
\texttt{new\_D\_omega\_flag = .true.}\\

for the \texttt{\&star\_job} section, and\\

\noindent\texttt{omega\_function\_weight = 20}\\

\noindent\texttt{D\_ST\_factor = 1}\\
\noindent\texttt{D\_visc\_factor = 0.0}\\

\noindent\texttt{am\_nu\_visc\_factor = 1.0}\\
\noindent\texttt{am\_nu\_ST\_factor = 1.0}\\
\texttt{am\_D\_mix\_factor = 0.0}\\
\texttt{am\_gradmu\_factor = 0.1d0}\\

for the \texttt{\&controls} section, respectively.

\begin{figure}
\center
\includegraphics[width=0.85\linewidth]{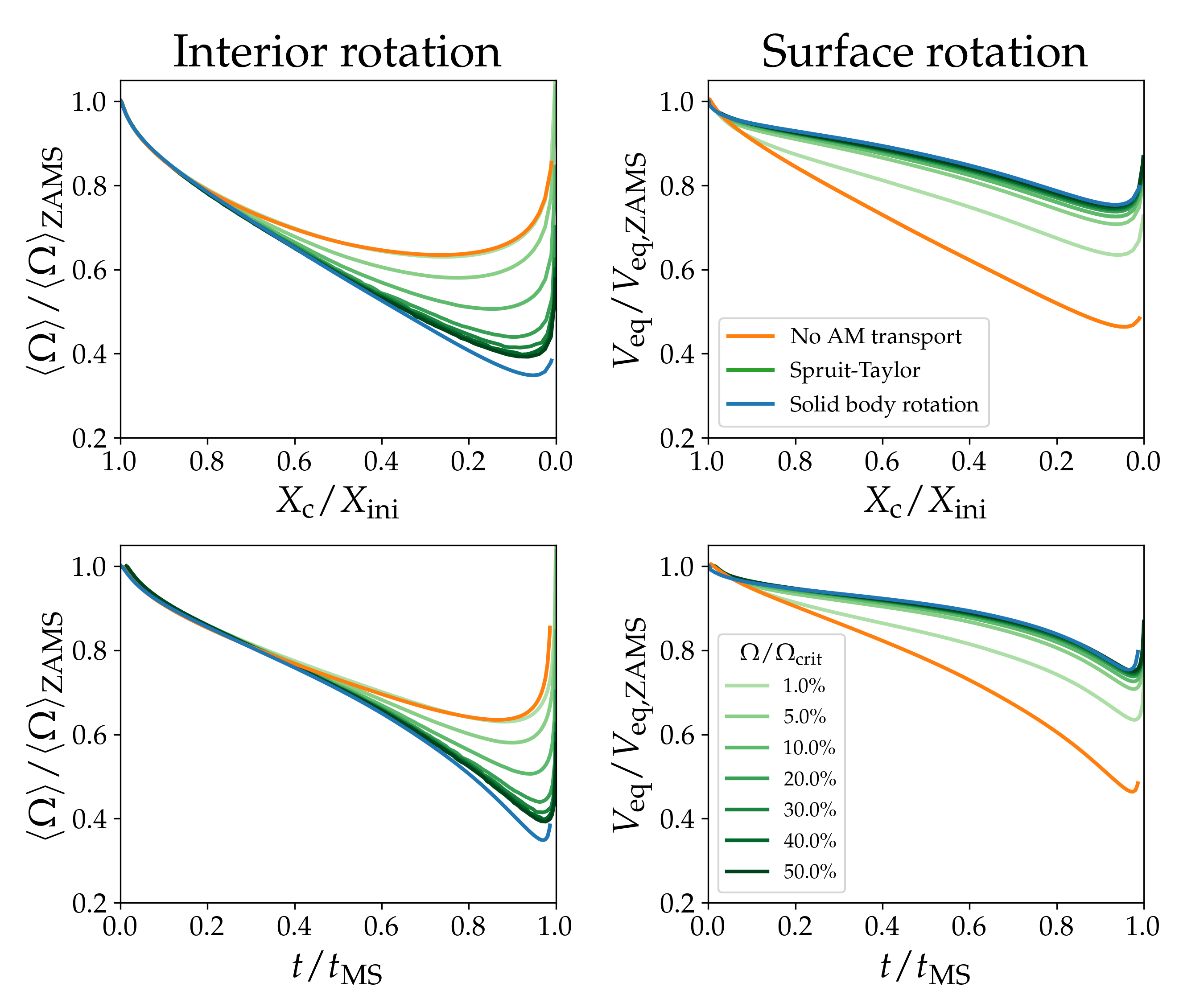}
\caption{Evolution of the average rotation frequencies (left) and the surface equatorial velocities (right) for a 5\,M$_\odot$ star as a function of the two age indicators $X_{\rm c}/X_{\rm ini}$ (top) and $t/t_{\rm MS}$ (bottom), assuming angular momentum transport from the Spruit-Taylor dynamo (green). The intensity of the green lines for the Spruit-Taylor dynamo depends on the initial rotation rate as indicated in the legend. The two cases of no angular momentum transport (orange) and solid body rotation (blue) are shown for comparison.}
	\label{fig:TS_vary_OmegaCrit}
\end{figure}

For comparison, we show in Fig.~\ref{fig:TS_vary_OmegaCrit} how varying the initial rotation rate from $1-50\%\, \Omega_{\rm crit}$ for a 5\,M$_\odot$ star changes the evolution of average rotation frequency experienced by the g-modes and the surface equatorial velocity, when angular momentum transport occurs through the Spruit-Taylor dynamo. As seen in the figure, the relative evolution starts to converge for $>30\%\, \Omega_{\rm crit}$ and approaches the curve for the solid body rotation.


\section{Amplitude corrections}\label{App:Amp_corr}

The pulsation properties of the sample of 52~SPB stars considered in this work were studied using a variety of different telescopes, see e.g. second column of Tabel~\ref{Tab:SPB_sample}, and therefore the recorded dominant pulsation amplitudes were obtained using different photometric passbands. Therefore, when comparing these amplitudes as done in Figs.~\ref{fig:inclination_space} and \ref{fig:inclination_ground} it is important to make sure that the differences in passbands have been accounted for. Furthermore, for the sample of SPB stars where the asteroseismic analysis was done using ground-based data, the amplitudes in the literature are given in milli-magnitudes (mmag) whereas they are given in parts-per-million (ppm) for the SPB stars with space photometry. As a first step, we convert all amplitudes in mmag to ppm using

\begin{equation}
{\rm Amplitude \ (ppm)} = 10^3 \times \frac{{\rm Amplitude \ (mmag)}}{\log_{10} e}.
\end{equation}

Next, we follow the methodology outlined by \cite{Lund2019} to convert all amplitudes to the Johnson V band. Here the conversion factors are obtained by first deriving the conversion factor\footnote{${\rm Amplitude}_{\rm bol} = C_{P - {\rm bol}}\times {\rm Amplitude}_P$} $C_{P - {\rm bol}}$ between the bolometric amplitudes (${\rm Amplitude}_{\rm bol}$) and the amplitudes ($ {\rm Amplitude}_P$) obtained for a given passband $P$ using Eq.~(2) in \cite{Lund2019} and assuming black-body radiation. The the amplitude conversion factor between the Johnson V band and a different passband $P$ is calculated as 

\begin{equation}
C_{P - V} = \frac{C_{P - {\rm bol}} (T_{\rm eff})}{C_{V - {\rm bol}}(T_{\rm eff})},
	\label{Eq:amp_corr}
\end{equation}

such that the amplitudes in the Johnson V band can be derived using

\begin{equation}
{\rm Amplitude}_V = C_{P - V} \times {\rm Amplitude}_P.
\end{equation}

\begin{figure}
\center
\includegraphics[width=0.75\linewidth]{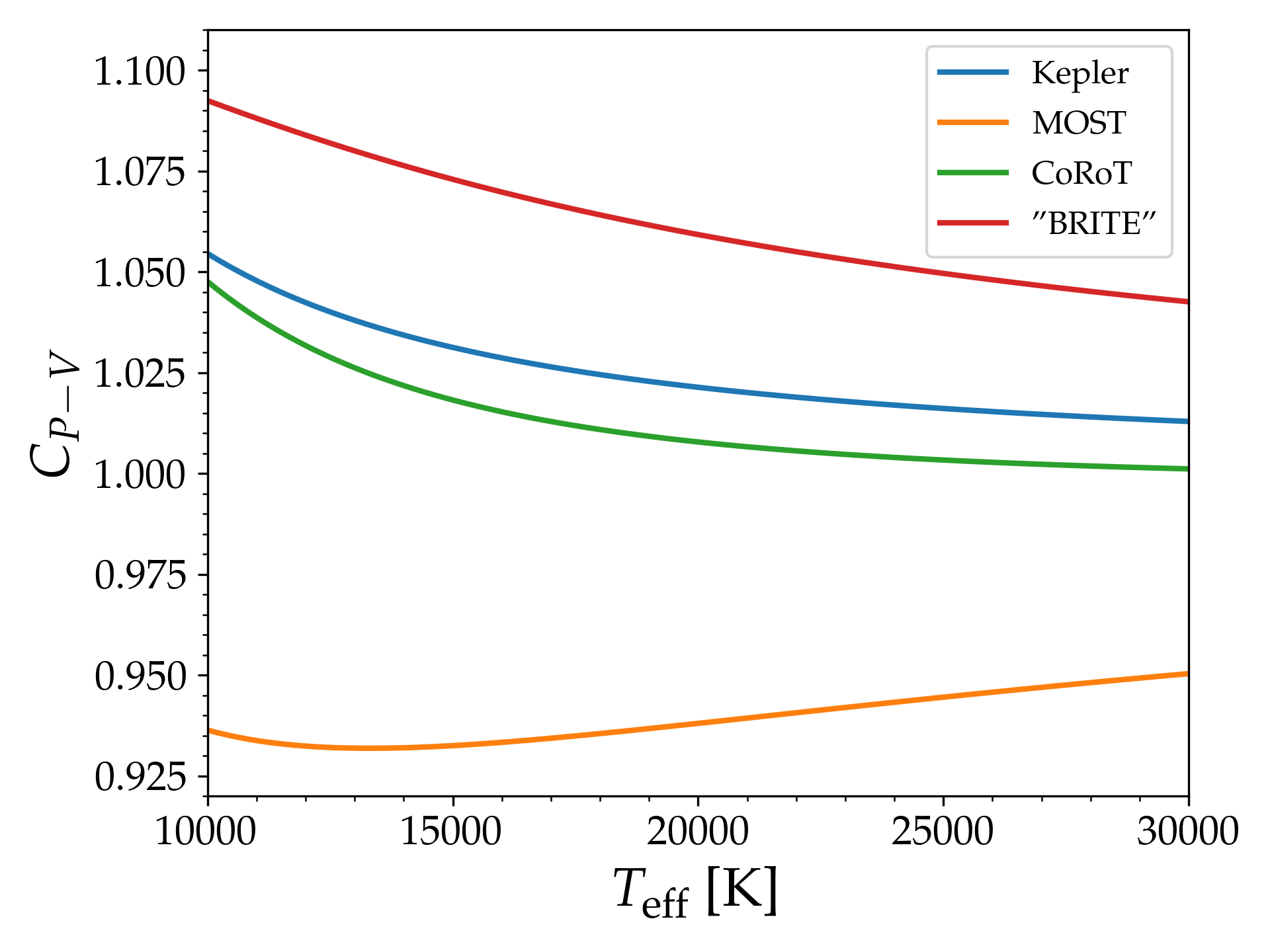}
\caption{Amplitude conversion factors derived using Eq.~\ref{Eq:amp_corr}  as a function of effective temperature.}
	\label{fig:amp_corr}
\end{figure}

The photometric passbands used for deriving the conversion factors between different passbands were obtained from the SVO filter service\footnote{\url{http://svo2.cab.inta-csic.es/theory/fps/}} \citep{Rodrigo2012,Rodrigo2020}. For the 19~SPB stars with ground-based data, the amplitudes are already given in the Johnson V band. Figure~\ref{fig:amp_corr} shows $C_{P - V}$ as a function of the effective temperature. For the photometric passband of the BRITE space telescope the SDSS.r photometric passband was used \citep{Weiss2014}. We note that the derived corrections are an estimate of the true $C_{P - V}$ values, which also depend on the metallicity, surface gravity, and the reddening of the stars.

\bibliography{references.bib}{}
\bibliographystyle{aasjournal}


\end{document}